\documentclass{article}

\usepackage[utf8]{inputenc}
\usepackage{graphicx}
\usepackage[breaklinks]{hyperref}

\usepackage{subfig}
\usepackage{hyperref}
\usepackage[framed,numbered,autolinebreaks,useliterate]{mcode}

\begin{document}
%\frontmatter
%\include{foreword}
%\tableofcontents
%\include{preface}
%\mainmatter
%\part{}
\def\figs{./figs/}
\def\labela{(\textbf{a}) }
\def\labelb{(\textbf{b}) }
\def\labelc{(\textbf{c}) }
\def\labeld{(\textbf{d}) }
\def\mycap#1#2#3{\caption{\label{#1}\textbf{#2 }#3}}
%
% ---------------------------------------------------------------
% IMPORTANT: I REMOVED PACKAGE 'stmaryrd.sty'  IN  'wileyvch.sty'
% ---------------------------------------------------------------
%
%\setcounter{chapter}{-1}	% my chapter will have number 12
\title{From Neuronal Models to Neuronal Dynamics and Image Processing
\footnote{First published as chapter 10 (pp.221-243) in the book:\textit{Biologically Inspired Computer Vision: Fundamentals and Applications}. Gabriel Cristobal, Laurent Perrinet, \& Matthias S. Keil (Editors),
ISBN: 978-3-527-68047-4, Wiley, 2015}}
\author{Matthias S. Keil\\
University of Barcelona, Faculty of Psychology, Campus Mundet\\Passeig de la Vall d'Hebron, 171, E-08035 Barcelona (Spain)\\EMail: matskeil[AT]ub.edu}
\maketitle
%\noindent
%
%
%=============================================================================
\section[Introduction]{Introduction}
%=============================================================================
%
% write at the end!!
% An introduction to the anatomy: dendrites - cell body etc
% Include a road-map for your chapter
%
%
Neurons make contacts with each other by synapses: The presynaptic neuron sends its output  to the synapse via an axon, and
the postsynaptic neuron receives it via its dendritic tree (``dendrite'').   Most neurons produce their output in form of
virtually binary events (action potentials or spikes).  Some classes of neurons (e.g., in the retina) nevertheless do not
generate spikes but show continuous responses.
Dendrites were classically considered as passive cables whose only function is to transmit input signals to the soma ($=$ cell
body of the neuron).  In this picture, a neuron integrates all input signals and generates a response if their sum exceeds 
a threshold.  Therefore, the neuron is the site where computations take place, and information is stored across the network in
synaptic weights (``connection strengths'').    This ``connectionist'' point of view on neuronal functioning inspired  neuronal
networks learning algorithms such as error backpropagation \cite{Backprop1986}, and more recent deep-learning architectures
\cite{DeepLearning2006}.   Recent evidence, however, suggest that dendrites are excitable structures rather than passive
cables, which can perform sophisticated computations \cite{MelReview94,HaeusserEtAl00,SegevLondon00,LondonHaeusserl05}.
This suggest that even single neurons can carry out far more complex computations than previously thought.\\
Naturally, a modeller has to choose a reasonable level of abstraction.  A good model is not necessarily one which incorporates all possible
details, because too many details can make it difficult to identify important mechanisms.   The level of detail is related to the research
question that we wish to answer, but also to our computational resources \cite{BretteEtAl2006}.  For example, if our interest was to
understanding the  neurophysiological details of a small circuit of neurons, then we probably would choose a Hodgkin-Huxley model
for each neuron \cite{HodkinHuxley1952} and also include detailed models of dendrites, axons, and synaptic dynamic and learning.
A disadvantage of the Hodgkin-Huxely model is its high computational complexity.  It requires about $1200$ floating point operations
(FLOPS) for the simulation of $1$ms of time \cite{Izhikevich2004}.\\
Therefore, if we want to simulate a large network with many neurons, then we may omit dendrites, axons and use a simpler model
such as the integrate-and-fire neuron model (which is presented further down) or the Izhikevich model \cite{Izhikevich2003}.
The Izhikevich model offers the same rich spiking dynamics as the full Hodgkin-Huxley model (e.g. bursting, tonic spiking, ...), while
having a computational complexity similar to the integrate-and-fire neuron model (about $13$ and $5$ FLOPS$/1$ms, respectively).
Further simplifications can be made if we aim at simulating psychophysical data, or at solving image processing tasks.  For example,
spiking mechanisms are often not necessary then, and we can compute a continuous response $y$ (as opposed to binary spikes)
from the neuron's state variable $x$ (e.g., membrane potential) by thresholding (half-wave rectification, e.g. $y=max[x,0]$) or via
a sigmoidal function (``squashing function'', e.g. $y=1/[1+\exp(-x)]$).\\
This chapter approaches neural modeling and computational neuroscience, respectively, in a tutorial-like fashion.    This
means that basic concepts are explained by way of simple examples, rather than providing an in-depth review on a specific
topic.  We proceed by first introducing a simple neuron model (the membrane potential equation), which is derived by
considering a neuron as an electrical circuit.  When the membrane potential equation is augmented with a spiking mechanism,
it is known as the leaky integrate-and-fire model.  ``Leaky'' means that the model forgets exponentially fast about past inputs.
``Integrate'' means that it sums its inputs, which can be excitatory (driving the neuron towards its response threshold) or
inhibitory (drawing the state variable away from the response threshold).  Finally, ``fire'' refers to the spiking mechanism.\\
The membrane potential equation is subsequently applied to three image processing tasks.  Section \ref{retina} presents a
reaction-diffusion-based model of the retina, which accounts for simple visual illusions and afterimages.
Section \ref{texture} describes a model for segregating texture features that is inspired by computations 
in the primary visual cortex \cite{MatsThesis}.  Finally, section \ref{collision} introduces a model of a
collision-sensitive neuron in the locust visual system \cite{MatsEliAngel04}.  The last section discusses
a few examples where image processing (or computer vision) methods also could explain how the brain processes
the corresponding information.
%
%--------------------------------------------------------------
\section{The Membrane Equation as a Neuron Model}
%--------------------------------------------------------------
%
\def\Na{$\mathrm{Na}^+\,$}
\def\K{$\mathrm{K}^+\,$}
\def\Cl{$\mathrm{C}l^-\,$}
\def\gleak{g_\mathrm{leak}}
\def\Vrest{V_\mathrm{rest}}
\def\Vreset{V_\mathrm{reset}}
\def\Vthresh{V_\mathrm{thresh}}
\def\Vexc{V_\mathrm{exc}}
\def\Vinh{V_\mathrm{inh}}
In order to derive a simple yet powerful neuron model, imagine the neuron's cell membrane (comprising dendrites, soma, and axon)
as a small sphere.   The membrane is a bilayer of lipids, which is $30$-$50\,$\AA$\,$  thick ($1\,$\AA$\,=10^{-10}$m$=0.1$nm). 
It isolates the extracellular space from a cell's interior, and thus forms a barrier for different ion species, such as \Na (sodium), \K (potassium), 
and \Cl (chloride).
Now if the ionic concentrations in the extracellular fluid and the cytoplasm ($=$ cell's interior) were all the same, then one would be probably
dead.   Neuronal signalling relies on the presence of ionic gradients.  With all cells at rest (i.e., neither input nor output), about half of the
brain's energy budget is consumed for moving \Na to the outside  of the cell and \K inward (the $\mathrm{Na}^+$ - $\mathrm{K}^+$-pump). 
Some types of neurons also pump \Cl outside (via the $\mathrm{Cl}^-$-transporter).   The pumping mechanisms compensate for the ions 
that leak through the cell membrane in the respective reverse directions, driven by electrochemical gradients.  At rest, the neuron
therefore maintains a dynamic equilibrium.\\
We are now ready to model the neuron as an electrical circuit, where a capacitance (cell membrane) is connected in parallel with a (serially
connected)  resistance and a battery.   The battery sets the cell's resting potential $\Vrest$.   In particular, all (neuron-, glia-, muscle-)
cells have a negative resting potential, typically $\Vrest=-65$mV.   The resting potential is the value of the membrane voltage $V_m$
when all ionic concentrations are in their dynamic equilibrium.  This is of course a simplification since we lump together the diffusion
potentials (such as $V_{\mathrm{Na}^+}$ or $V_\mathrm{K}^+$) of each ion species.  The simplification comes at a cost, however, 
and our resulting neuron model will not be able to produce action potentials or spikes by itself (the binary events by which many
neurons communicate with each other) without explicitly adding a spike-generating mechanism (more on that in section \ref{Spikes}).
By Ohm's law, the current that leaks through the membrane is $I_\mathrm{leak}=\gleak (V_m-Vrest)$, where the \emph{leakage
conductance} $\gleak$ is just the inverse of the membrane resistance.  
The charge $Q$ which is kept apart by a cell's membrane with capacitance $C$ is $Q(t)=CV_m(t)$.   Whenever the neuron signals, the
distribution of charges changes, and so does the membrane potential, and thus $dQ(t)/dt\equiv\, I_C=CdV_m(t)/dt$ will be non-zero. 
In other words, a current $I_C$ will flow, carried by ions.  Assuming some fixed value for $I_C$,  the change in $V_m$ will be slower
for a higher capacitance $C$ (better buffering capacity). \emph{Kirchhoff's current law} is the equivalent of current conservation
$I=I_C+I_\mathrm{leak}$, or
\begin{equation}\label{MembraneEquation}
   C\frac{dV_m(t)}{dt} + \gleak(V_m-\Vrest)=I
\end{equation}
The right hand side corresponds to current flows due to excitation and inhibition (more on that later).  Biologically,  current flows occur
across protein molecules that are embedded in the cell membrane.  The various protein types implement specific functions such as ionic
channels, enzymes, pumps, and receptors.
These ``gates'' or ``doors'' through the cell membrane are highly specific, such that only particular information or substances (like ions) 
can enter or exit the cell.   Strictly speaking, each channel which is embedded in a neuron's cell membrane would corresponds to a RC-circuit
($R=1/\gleak=$ resistance, $C=$capacitance) such as equation \ref{MembraneEquation}.   But fortunately, neurons are very small, what
justifies the assumption that channels are uniformly distributed, and  the potential does not vary across the cell membrane:  The cell is said
to be \emph{isopotential}, and it can be adequately described by a \emph{single} RC-compartment.\\
Let's assume that we have nothing better to do than waiting for a sufficiently long time such that the neuron reaches equilibrium. 
Then, by definition, $V_m$ is constant, and thus $I_C=0$.  What remains from the last equation is just $(V_m-\Vrest)/R=I$, or
$V_m=\Vrest + RI$.  In the absence of excitation and inhibition we have $I=0$ and the neuron will be at its resting potential. 
But how long do we have to wait until the equilibrium is reached?  To find that out, we just move all terms with $V_m$ to one side
of the equation, and the term which contains time to the other.  This technique is known as \emph{separation of variables}, and
permits integration in order to convert the infinitesimal quantities $dt$ and $dV_m$ into "normal" variables (we formally rename 
the corresponding integration variables for time and voltage as $T$ and $V$, respectively),
\begin{equation}
  \int_{V_0}^{V_m}\frac{dV}{V-\Vrest-RI} = -\frac{1}{RC}\int_{0}^{t} dT
\end{equation}
where $V_0\equiv V_m(t=0)$.   We furthermore define $V_{\infty}\equiv RI$.  Integration of the left hand side gives
$\log[(V_m-\Vrest-V_{\infty})/(V_0-\Vrest-V_{\infty})]$.  With the (membrane) time constant of the cell $\tau\equiv RC$, the above
equation yields
\begin{equation}
  V_m(t)=\left(V_0-\Vrest-V_{\infty}\right) e^{-t/\tau} + \Vrest + V_{\infty}
\end{equation}
It is easy to see that for $t\rightarrow \infty$ we get $V_m= \Vrest + V_{\infty}$, where $V_{\infty}=0$ in the absence of external
currents $I$ (this confirms our previous result).   The time that we have to wait until this equilibrium is reached depends on $\tau$:
The higher the resistance $R$, and the bigger the capacitance $C$, the longer it will take (and vice versa).\\
The constant $V_0$ has to be selected according to the initial conditions of the problem.  For example, if we assume that the
neuron is at rest when we start the simulation, then $V_0=\Vrest$, and therefore
\begin{equation}
  V_m(t)=V_{\infty} \left(1-e^{-t/\tau} \right) + \Vrest
\end{equation}
%
%
%
%
%--------------------------------------------------------------
\subsection{\label{SynapticInputs}Synaptic Inputs}
%--------------------------------------------------------------
%
Neurons are not loners but are massively connected to other neurons.  The connection sites are called \emph{synapses}.   Synapses come in two
flavours: Electrical and chemical.  Electrical synapses (also called \emph{gap junctions}) can directly couple the membrane potential of neighbouring neurons.
In this way, distinct networks of specific neurons are formed, usually between neurons of the same type.  Examples of electrically coupled neurons
are retinal horizontal cells \cite{Kolb77,NelsonEtAl85}, cortical low-threshold-spiking interneurons, and cortical fast-spiking interneurons
\cite{GalHes99,GiBeCo99} (interneurons are inhibitory).   Sometimes chemical and electrical synapses even combine to permit reliable and
fast signal transmission, such as it is the case in the locust, where the Lobula Giant Movement Detector (LGMD) connects to the Descending
Contralateral Movement Detector (DMCD) \cite{OSheaRowell1975,Rind84,KillmannSchuermann1985,KillmannGrasSchuermann1999}.\\
Chemical synapses are far more common than gap junctions.  In one cubic millimeter of cortical grey matter there are about one billion ($\approx 10^9$)
chemical synapses (ca. $10^{15}$ in the whole human brain).  Synapses are usually plastic.  Whether they increase or decrease their connection
strength to a post-synaptic neuron depends on causality.  If a pre-synaptic neuron fires within some $5-40$ms before the post-synaptic neuron,
the connection gets stronger (potentiation: "P")  \cite{MarkramEtAl1997}.  In contrast, if the pre-synaptic spike arrives
after activation of the post-synaptic neuron, synaptic strength is decreased (depression: "D") \cite{WaGeNa2005}.  This mechanism is known as
\emph{spike time dependent plasticity} (STDP), and can be identified with Hebbian learning \cite{Hebb49,SongMillerAbbott2000}. 
Synaptic potentiation is thought to be triggered by back-propagating calcium spikes in the dendrites of post-synaptic neuron \cite{MageeJohnston1997}
Synaptic plasticity can occur over several timescales, short term (ST) and long term (LT).   Remember these acronyms if you see letter
combinations like ``LTD'', ``LTP'', ``STP'' or ``STD''.\\
An activation of fast, chemical synapses causes a rapid and transient voltage change in the post-synaptic neuron.  These voltage changes
are called \emph{post-synaptic potentials} (PSPs).  PSPs can be either inhibitory (IPSPs) or excitatory (EPSPs).   Excitatory neurons 
depolarize their target neurons ($V_m$ will get more positive as a consequence of the EPSP), whereas inhibitory neurons hyperpolarize
their post-synaptic targets.   How can we model synapses?  PSPs are caused by a temporary increase in membrane conductance in series
with a so-called \emph{synaptic reversal battery} $E$ (also synaptic reversal potential, or synaptic battery).  The synaptic input
defines the current on the right hand side of equation \ref{MembraneEquation},
\begin{equation}\label{ExcitationInhibition}
  I=\sum_{i=1}^N g_i(t)\cdot (E_i-V_m)
\end{equation}
The last equation sums $N$ synaptic inputs, each with conductance $g_i$ and reversal potential $E_i$.  Notice that whether an input
$g_i$ acts excitatory or inhibitory on the membrane potential $V_m$ depends usually on whether the synaptic battery $E_i$ is bigger
or smaller than the resting potential $\Vrest$.  Just consider only one type of excitatory and inhibitory input.  Then we can write
\begin{equation}\label{MembraneEquation2}
   C\frac{dV_m(t)}{dt} = \gleak(\Vrest-V_m)+ g_{exc}(t)\cdot (\Vexc-V_m) + g_{inh}(t)\cdot (\Vinh-V_m)
\end{equation}
(For all simulations, if not otherwise stated, we assume $C=1$ and omit the physical units).
How to solve this equation? After converting the differential equation into a difference equation, the equation can be solved
numerically with standard integration schemes, such as Euler's method, Runge-Kutta, Crank-Nicolson, or Adams-Bashforth (see
for example  chapter 6 in \cite{Strang2007} and chapter 17 in \cite{NumericalRecipes3rd} for more details).  Typically, model
neurons are integrated with a step size of $1$ms or less, as this is the relevant time scale for neuronal signalling.  If the
simulation focusses more on perceptual dynamics (or biologically-inspired image processing tasks), then one may choose a bigger
integration time constant as well.  The ideal integration method is stable, produces solutions with a high accuracy, and has a
low computational complexity.  In practice, of course, we have to make the one or the other trade-off.\\
Remember that $\gleak\equiv 1/R$ is called \emph{leakage conductance}, which is just the inverse of the membrane resistance. 
For constant capacitance $C$, the leakage conductance determines the time constant $\tau\equiv C/\gleak$ of the neuron: 
Bigger values of $\gleak$ will make it "faster" (i.e. less memory on past inputs), while smaller values will cause a higher
degree of  low-pass filtering of the input.  $\Vexc > \Vrest$ and $\Vinh\leq \Vrest$ is the excitatory and inhibitory
synaptic battery, respectively.\\ 
For $\Vexc > \Vrest$ the synaptic current (mainly $Na^+$ and $K^+$) is inward and negative by convention. 
The membrane thus gets depolarized.  This is a signature of an EPSP.  In the brain, the most common type of excitatory
synapses release \emph{glutamate} (a neurotransmitter)\footnote{In the peripheral nervous system of vertebrates, excitatory
synapses are activated instead by acetylcholine (ACh)}.  The neurotransmitter diffuses across the synaptic cleft, and
binds on glutamate-sensitive receptors in the post-synaptic cell membrane.  As a consequence, ion channels will open,
and $Na^+$ and $K^+$ (but also $Ca^{2+}$ via voltage sensitive channels) will enter the cell.\\
\emph{Agonists} are pharmacological substances that do not exist in the brain, but open these channels as well.
For instance, the agonist NMDA\footnote{$N$-methyl-D-aspartat}  will open  excitatory,  voltage-sensitive NMDA-channels. 
AMPA\footnote{$\alpha$-amino-3-hydroxy-5-methyl-4-isoxalone  propionic acid} is another agonist that
activates fast excitatory synapses.   However, AMPA-synapses will remain silent in the presence of NMDA, and vice
versa.   Therefore one can imagine the ionic channels  as locked doors.   For their opening, the right key is necessary,
which is either a specific neurotransmitter, or some ``artificial'' pharmacological agonist.  The ``locks'' are the receptor sites
to which a neurotransmitter or an agonist binds.  The reversal potentials of the fast AMPA-synapse is about $80-100$mV
above the resting potential.  Usually, AMPA-channels co-occur with NMDA-channels, what may enhance the computational
power of a neuron \cite{Mel93}. \\
For $\Vinh<\Vrest$ the membrane is hyperpolarized.  For $\Vinh\approx \Vrest$, it gets less sensitive to depolarization,
and accelerates the return to $\Vrest$ for any synaptic input.
Pre-synaptic release of GABA\footnote{From ref. \cite{JonBuz07}: $\gamma$-aminobutryc acid} can activate three subtypes of receptors:
GABA$_A$, GABA$_B$, and GABA$_C$ (as before, they are identified through the action of specific pharmalogicals)\footnote{GABA$_A$, 
are ligand-gated ion channels permeable to $Cl^-$.   Post-synaptic GABA$_B$ are heptahelical receptors coupled to inwardly rectifying
$K^+$ channels.   Finally, GABA$_C$ are ligand-gated $Cl^-$ channels, which are primarily expressed in the retina.}\\
Why are the synaptic batteries also called reversal potentials? For excitatory input, $\Vexc$ imposes an upper limit on $V_m$. 
This means that no matter how big $g_{exc}$ will be, it can drive the neuron only up to $\Vexc$.   In order to understand that,
consider $(\Vexc-V_m)$, the so-called \emph{driving potential}:  If $V_m<<\Vexc$, then the driving potential is high, and the
neuron depolarizes fast.  The closer $V_m$ gets to $\Vexc$, the smaller the driving potential, until the excitatory current
$g_{exc}(t)\cdot (\Vexc-V_m)$ eventually approaches zero.   (Analog considerations hold for the inhibitory input).\\
What is the value of $V_m$ at equilibrium?  Equilibrium means that $V_m(t)$ does not change with $t$, and then the
left hand side of equation \ref{MembraneEquation2} is zero.  Of course this implies that all excitatory and inhibitory
inputs vary sufficiently slow and we can consider them as being constant.  Or, otherwise expressed, the neuron reaches
the equilibrium before a significant change in  $g_{exc}(t)$ or  $g_{inh}(t)$ takes place.  Then, solving equation
\ref{MembraneEquation2} for $V_m$ yields
\begin{equation}\label{MembraneEquationSteadyState}
V_m(t\rightarrow \infty) = \frac{\gleak\Vrest+ g_{exc}\Vexc + g_{inh} \Vinh}{\gleak+ g_{exc}+ g_{inh}}
\end{equation}
The time until the equilibrium is reached depends not only on $\gleak$, but on all other active conductances.
As a consequence, a neuron which receives continuously input from other neurons can react faster than a
neuron which starts from $\Vrest$ \cite{VreeswijkSompolinsky1996}.   (Ongoing cortical activity is the
normal situation, where it is thought that excitation and inhibition are just balanced \cite{HaiderEtAl2006}).\\
A specially interesting case is defined by $\Vinh=\Vrest$, which is called silent or \emph{shunting inhibition}.
It is silent because it only gets evident if the neuron is depolarized (and hyperpolarized if more than one type
of inhibition is considered).   Shunting inhibition decreases the time constant of the neuron, thus making it faster.  
In this way, the return to the resting potential
is accelerated, for excitatory and inhibitory input.  Furthermore, \emph{divisive inhibition} is a special form of
shunting inhibition if $\Vrest=0$.   With spiking neurons, however, pure divisive inhibition does not seem to
exist.  In that case, shunting inhibition is rather subtractive \cite{HoltKoch97} and cannot act as a gain control
mechanism.   But in networks with balanced excitation and inhibition, the choice is ours':  If we change the
balance between excitation and inhibition, then the effect on a neuron's response will be additive and subtractive,
respectively.  If we leave the balance unchanged and increase or decrease excitation and inhibition in parallel,
then a multiplicative or divisive effect on a neuron's response will occur \cite{AbbottChance2005}.\\
\begin{figure}[th!]
 \centering
  \subfloat[postsynaptic potentials]{\includegraphics[width=0.5\linewidth]{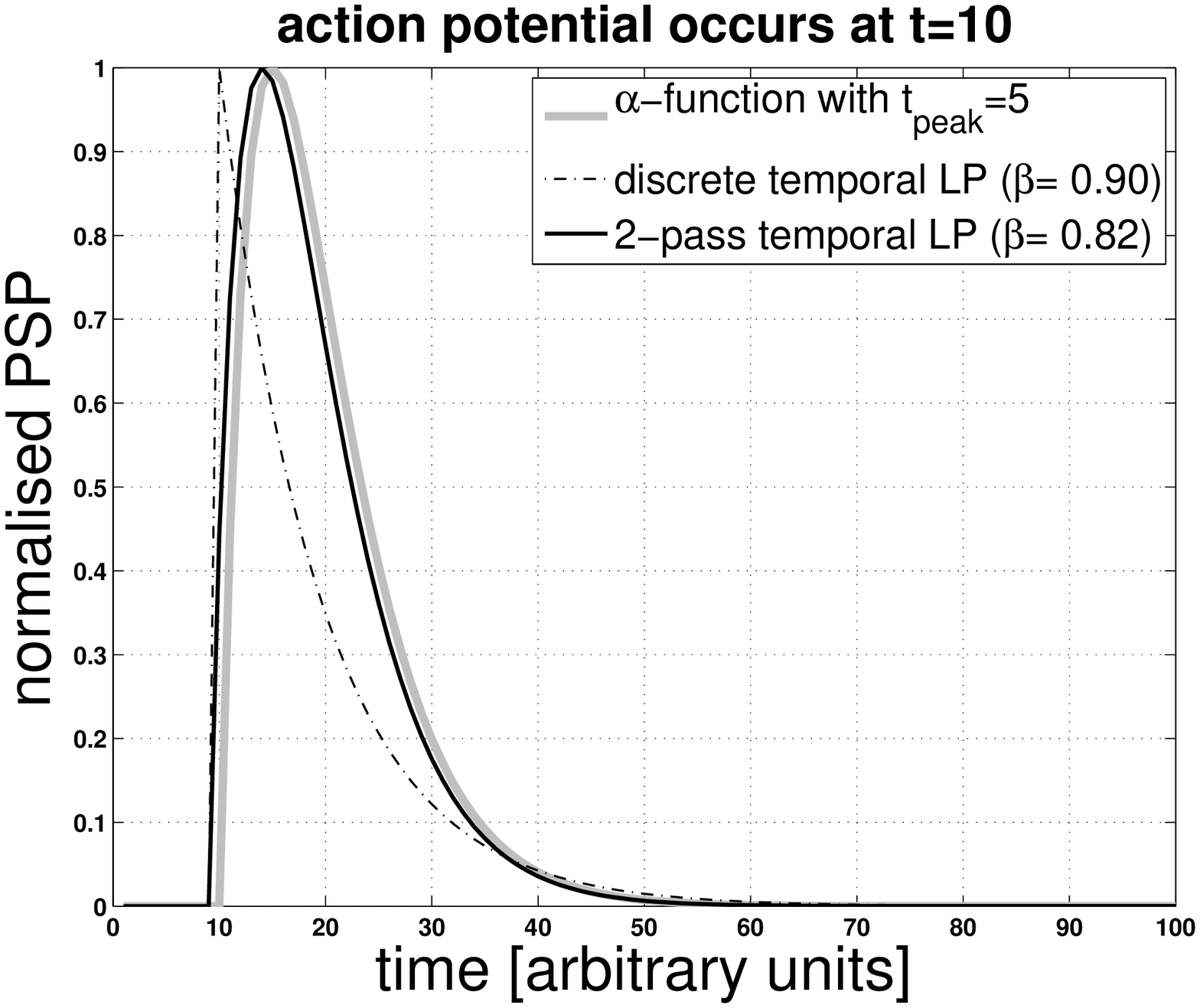}}
  \subfloat[actions potentials (APs or spikes)]{\includegraphics[width=0.5\linewidth]{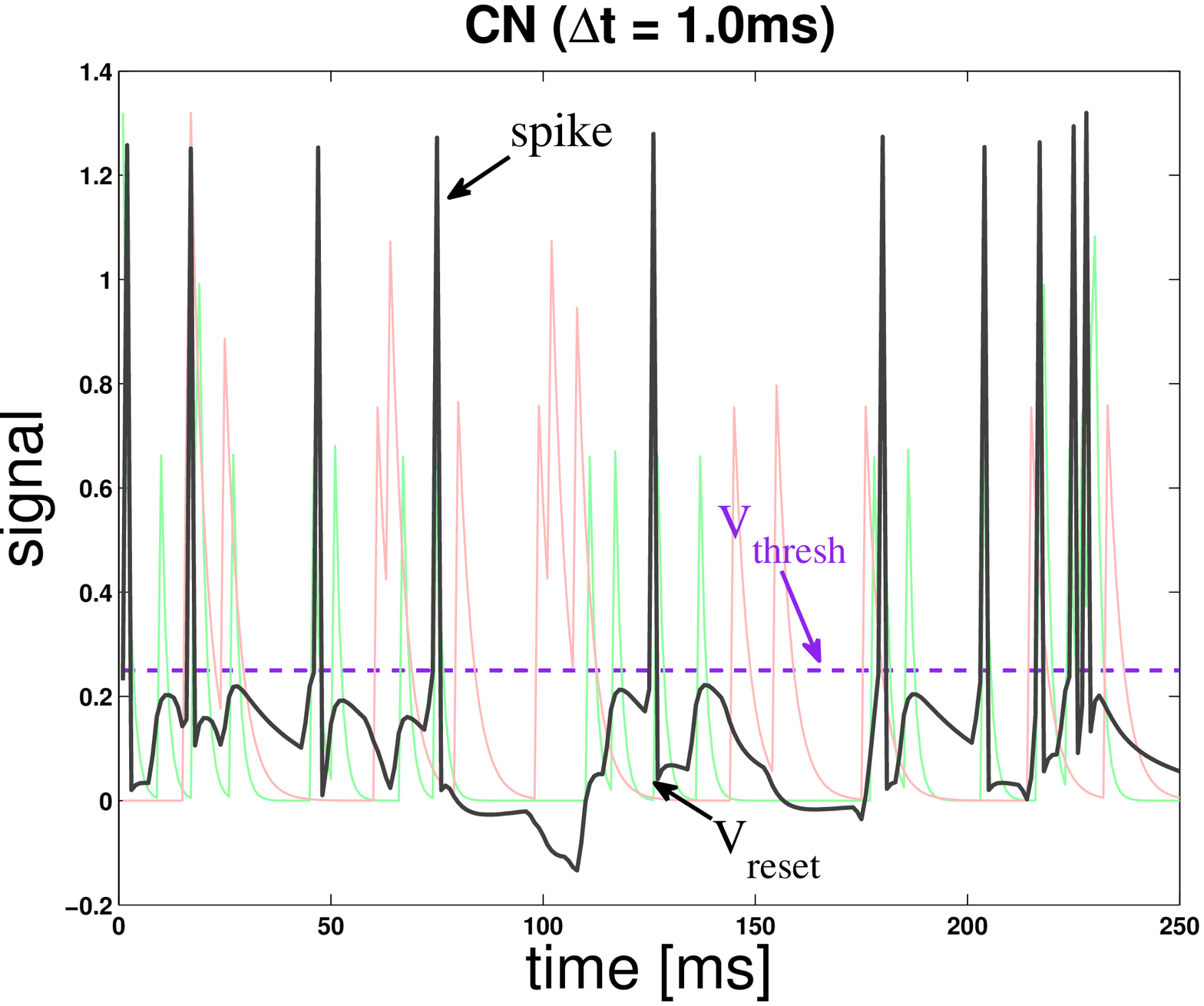}}
\mycap{SpikesAndPSPs}{Spikes and postsynaptic potentials}{\labela The figure shows three methods for converting a spike
(value one at $t_0=10$, that is $\delta[t-10]$) into a postsynaptic
potential (PSP).  The $\alpha$-function with $t_{peak}=5$ (equation \ref{ModifiedAlpha}) is
represented by the gray curve.  The result of low-pass filtering the spike once (via equation \ref{LPFilterPSP} with $\beta=0.9$)
is shown by the dashed line:  The curve has a sudden rise and a gradual decay.  Finally, applying two times the low-pass filter
(each with $\beta=0.82$) to the spike results in the black curve.  Thus, a 2-pass low-pass filter can approximate the
$\alpha$-function reasonably well.
\labelb The figure shows PSPs and output of the model neuron equation \ref{MembraneEquation2}
endowed with a spike mechanism.  Excitatory ($g_{exc}(t)$,  pale green curve) and inhibitory ($g_{inh}(t)$, pale red curve)
PSPs cause corresponding fluctuations in the membrane potential $V_m$ (black curve).  As soon as the membrane potential crosses
the threshold $\Vthresh=0.25$ (dashed horizontal line), a spike is added to $V_m$, and afterwards the membrane potential is reset
to $\Vreset=0$.  The half-wave rectified membrane potential represents the neuron's output.  The input to the model neuron
were random spikes which were generated according to a Poisson process (with rates $\mu_{exc}=100$ and $\mu_{inh}=50$ spikes
per second, respectively).  The random spikes were converted into PSPs via simple low-pass filtering
(equation \ref{LPFilterPSP}) with filter memories $\beta_{exc}=0.5$ and $\beta_{inh}=0.75$, respectively, and weight $w_{syn}=75$.
Integration method was Crank (not Jack)-Nicolson with step size $\Delta t=1$ms.  The rest of the parameters of equation
\ref{MembraneEquation2} were $\Vexc=3$, $\Vinh=-1$, $\Vrest=0$, and $\gleak=50$.}
\end{figure}%

%
%
%-------------------------------------------------------------------
\subsection{\label{Spikes}Firing Spikes}
%------------------------------------------------------------------
%
Equation \ref{MembraneEquation2} represents the membrane potential of a neuron, but $V_m$ could represent different quantities as well.
For example, $V_m$ could be interpreted directly as response probability if we set, for example, $\Vrest=0$, $\Vexc=1$, and $\Vinh=0$.
Accordingly, in the latter case we have $0 \leq V_m \leq 1$.   Another possibility is to set $\Vinh=-1$.   As neuronal responses are only positive,
however, $V_m$ has to be \textit{half-wave rectified}, meaning that we take $[V_m]^+ \equiv \max(0,V_m)$ as the output of the neuron\footnote{In
this case, the \emph{response threshold} $V_{thresh}=0$.   For arbitrary thresholds $\max(0, V_{thresh})$.}.  Naturally, half-wave rectification makes
only sense if negative values of $V_m$ can occur.  Because of the absence of explicit spiking, the neuron's output represents a (mean) \textit{firing rate},
usually interpreted as spikes per second. \\
When should one use equation \ref{MembraneEquation2} or \ref{MembraneEquationSteadyState}? This depends
mainly on the purpose of the simulation.  When the synaptic input consists of spikes, then one needs some
mechanism to convert them into continuous quantities.   The low-pass filtering characteristics of equation
\ref{MembraneEquation2} will do that.   For instance, spikes are necessary for implementing spike-time-dependent
plasticity (STDP),  which modifies synaptic strength dependent on pre- and postsynaptic activity.
For some purposes (e.g. biologically-inspired image processing),  spikes are not strictly necessary, and one can use
the (mean or instantaneous) firing rate (or activity, that is the rectified membrane potential) of pre-synaptic
neurons directly as input to post-synaptic neurons.   In that case, both equation \ref{MembraneEquation2}
or \ref{MembraneEquationSteadyState} could be used.   The steady-state solution (equation \ref{MembraneEquationSteadyState}),
however, has less computational complexity, as one does not need to integrate it numerically.   Recall that when using the steady-state
solution, one implicitly assumes that the synaptic input varies on a relatively slow time scale, such that the neuron can reach its
equilibrium state in each moment.\\
It is straightforward to convert \ref{MembraneEquation2} into a \emph{leaky integrate-and-fire neuron}.  
The term ``leaky'' refers to the leakage conductance $\gleak$, and the neuron would only be a perfect integrator for $\gleak=0$
As soon as the membrane potential $V_m(t)$ crosses the neuron's response threshold $V_{thresh}$, we record a pulse with some
amplitude in the neuron's response.    Otherwise the response is usually defined as being zero:
\vspace*{-3mm}
\begin{lstlisting}
	if V>Vthresh	% Vthresh = response threshold
		% response is a pulse with amplitude 'SpikeAmp'
		response = SpikeAmp;
		% reset membrane potential (afterhyperpolarization)
		V = Vreset;
	else
		response	= 0;	% else, the neuron stays silent
	end
\end{lstlisting}
In order to account for afterhyperpolarization (i.e., the refractory period), $V_m$ is set to some value $\Vreset$ after each spike.  Usually,
$\Vreset \leq \Vrest$ is chosen.  The refractory period of a neuron is the time that the ionic pumps need to re-establish the original
ion charge distributions.  Within the absolute refractory period, the neuron will not spike at all, or spiking probability will be greatly reduced 
(relative refractory period). 
Other spiking mechanisms are conceivable  as well.  For example, we can define the model neuron's response as being identical to firing rate
$[V_m]^+$, and add a spike to $V_m$ whenever $V_m>V_{thresh}$.  Then $V_{thresh}$ would represent a \emph{spiking threshold}:
\vspace*{-3mm}
\begin{lstlisting}
	if V>Vthresh	% 'Vthresh' = spiking threshold
		% add a spike with amplitude 'SpikeAmp' to current membrane potential 'V'
		response  = V + SpikeAmp;
		% reset membrane potential (refractory period)
		V = Vreset;
	else
		response = max(V,0);	% otherwise, rate-code-like response (by half-wave rectification of 'V')
	end
\end{lstlisting}
The response thus switches between a rate code ($V_m<V_{thresh}$) and a spike code  ($V_m\geq V_{thresh}$)  \cite{HuxterBurgessOKeefe2003,AlleGeiger2006}.
A typical spike train produced by the latter mechanism is shown in figure \ref{SpikesAndPSPs}b.\\
How are binary events such as spikes converted into post-synaptic potentials (PSPs)? A PSP has a sharp rise and a smooth decay,
and therefore is usually broader than the spike by which it was  evoked.  Assume that a spike arrives at time $t_i$ at the
post-synaptic neuron.  Then the time course of the corresponding PSP (i.e. the excitatory or inhibitory input to the neuron)
is adequately described by the so-called $\mathit{\alpha}$\textit{-function}:
\begin{equation}\label{ModifiedAlpha}
  \alpha(t,t_i)=\mathrm{const}\cdot (t-t_i) e^{- (t-t_i)/t_{peak}}\Theta(t-t_i)
\end{equation}
The constant is chosen such that $g_{syn}(t=t_{peak})$ matches the desired maximum of the PSP. 
The Heaviside function $\Theta(x)$ is is zero for $x \leq 0$, and $1$ otherwise.  It makes sure
that the PSP generated by the spike at $t_i$ starts at time $t_i$.  The total synaptic input $g_{syn}(t)$ 
into the neuron is $\sum_{i} \alpha(t,t_i)$, multiplied with a synaptic weight $w_{syn}$.\\
Instead of using the $\alpha$-function, we can simplify matters (and accelerate our simulation) by assuming that each spike
causes an instantaneous increase in $g_{exc}(t)$ or $g_{inh}(t)$, respectively, followed by an exponential decay.  Doing so 
just amounts to adding one simple differential equation to our model neuron:
\begin{equation}\label{DEQPSP}
 \tau \frac{dx}{dt}  = - x + w_{syn} \cdot \delta[t-t_i]
\end{equation}
The time constant $\tau$ determines the rate of the exponential decay (faster decay if smaller), $w_{syn}$ is the synaptic
weight, and $\delta[\cdot]$ is the Kronecker Delta function, which is just one if its argument is zero: The $i$-th spike
increments $x$ by $w_{syn}$.  The last equation low-pass filters the spikes.  An easy-to-compute discrete version 
can be obtained by converting the last equation into a finite difference equation, either by forward or backward differencing
(details can be found in section S8 of ref. \cite{MatsJoan2012}):
\begin{equation}\label{LPFilterPSP}
 x_{n+1}   =  \beta x_n + w_{syn}(1-\beta)\delta[t-t_i]
\end{equation}
For forward differentiation, $\beta=1-\Delta t/\tau$, where $\Delta t$ is the integration time constant that comes from approximating
$dx/dt$ by $(x_{n+1}-x_n)/\Delta t$.   For backward differencing, $\beta=\tau/(\tau+\Delta t)$.  The degree of low-pass filtering
($=$ filter memory on past inputs) is determined by $\beta$.  For $\beta=0$, the filter output $x_{n+1}$ just reproduces the input
spike pattern (the filter is said to have no memory on past inputs).  For $\beta=1$, the filter ignores any input spikes, and stays
forever at the value with which it was initialized (``infinite memory'').  For any value between zero and one, filtering takes place,
where filtering gets stronger with increasing $\beta$.\\
When one (or more) spikes are filtered by equation \ref{LPFilterPSP}, then we see a sudden increase in $x$ at time $t_i$, followed by
a gradual decay (figure \ref{SpikesAndPSPs}a, dashed curve).  This sudden increase stands in contrast to the gradual increase of the
PSP as predicted by equation \ref{ModifiedAlpha}  (figure \ref{SpikesAndPSPs}a, gray curve).  A better approximation to the shape of
a PSP results from applying low-pass filtering twice, as shown by the black curve in figure  \ref{SpikesAndPSPs}a.  This is tantamount
to simulating two equations \ref{LPFilterPSP} for each synaptic input to equation \ref{MembraneEquation2}.
\begin{figure}[th!]
 \centering
  \subfloat[grating induction]{\includegraphics[width=0.23\linewidth]{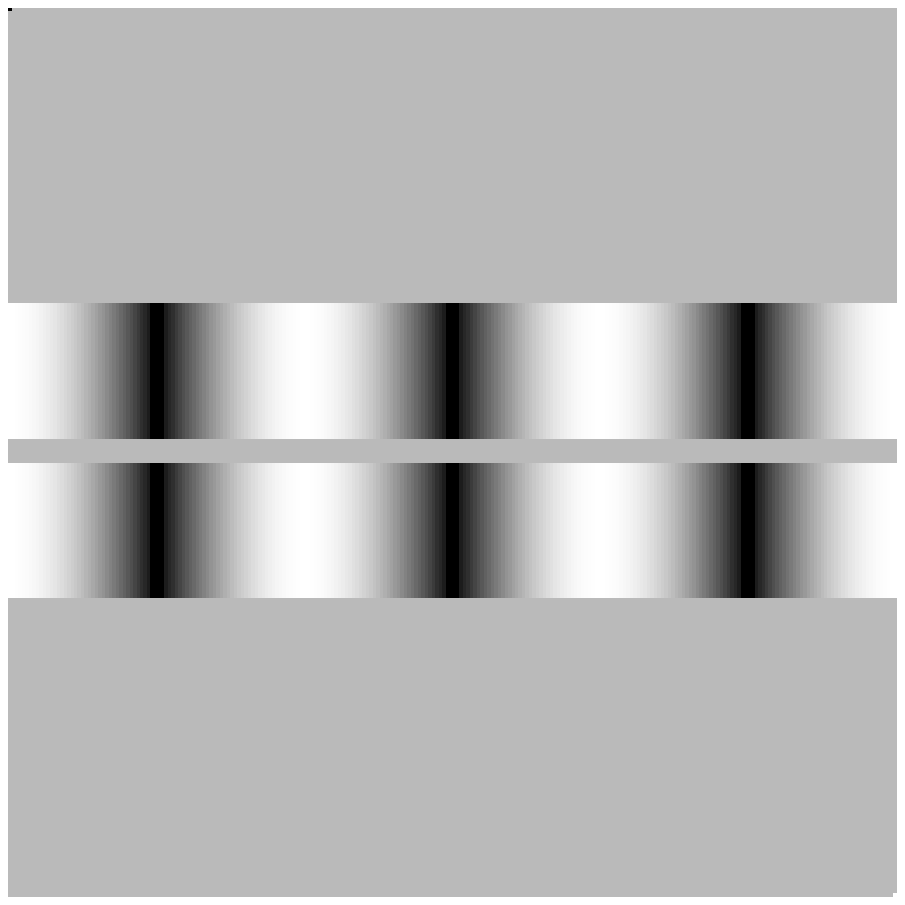}}~~
  \subfloat[no induction]{\includegraphics[width=0.23\linewidth]{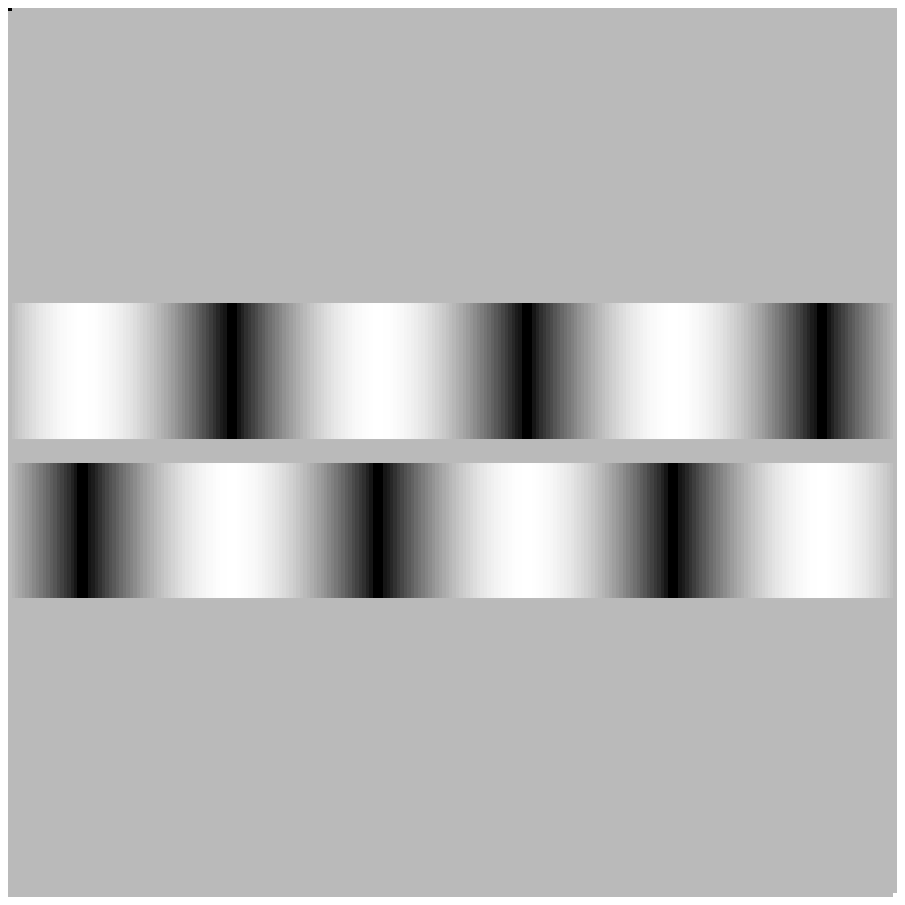}}~~
  \subfloat[camera]{\includegraphics[width=0.23\linewidth]{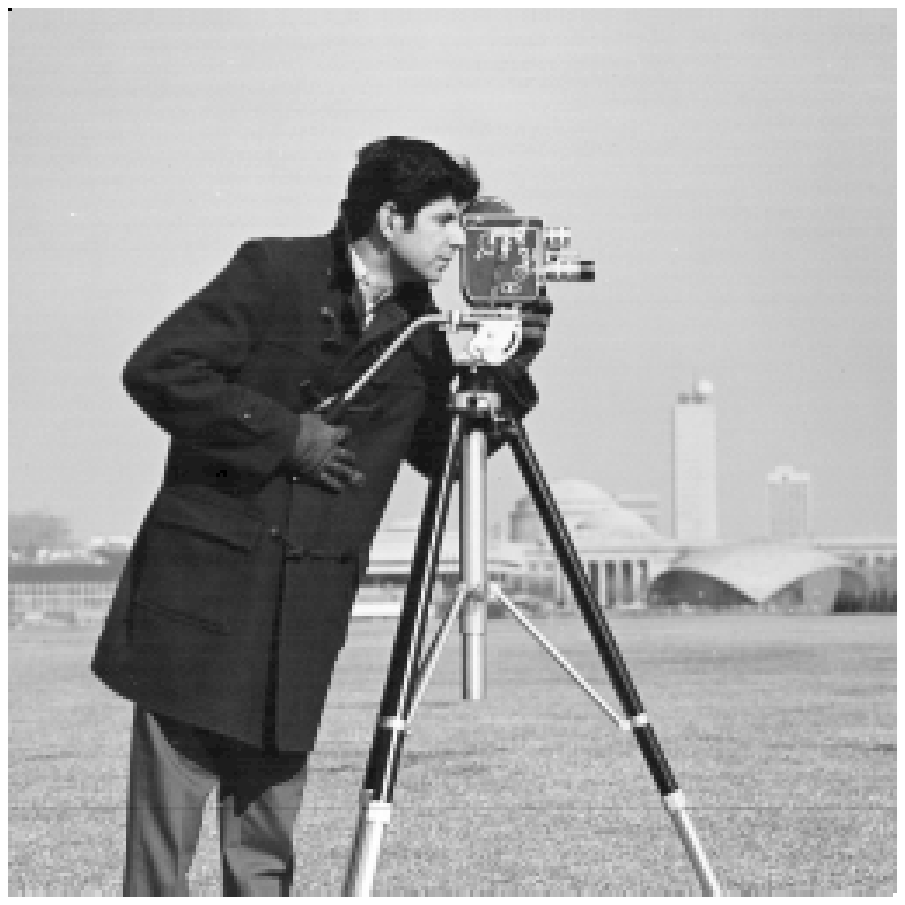}}~~
  \subfloat[staircase]{\includegraphics[width=0.23\linewidth]{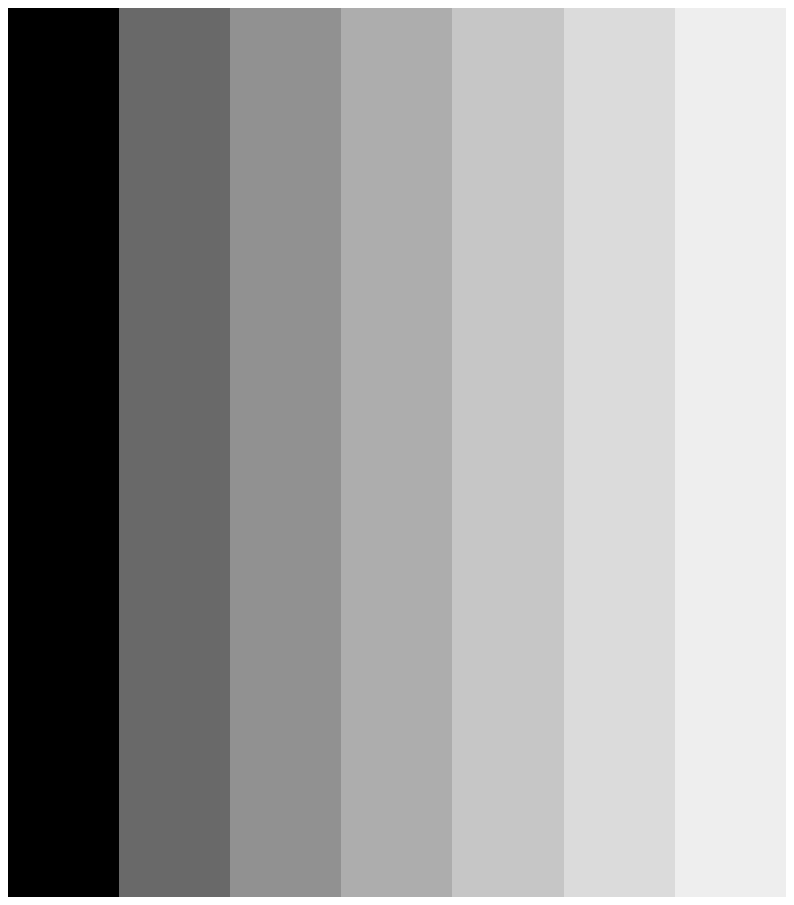}}
\mycap{Luminance}{Test images}{All images have $256$ rows and $256$ columns.
\labela The upper and the lower grating (`ìnducers'') are separated by a small
stripe, which is called the \emph{test stripe}.  Although the test stripe has the same luminance throughout, humans perceive
a wave-like pattern with opposite brightness than the inducers.  That is, where the inducers are white, the test stripe appears
darker and vice versa.
\labelb When the inducer gratings have an opposite phase (that is, white stands vis-a-vis black), then the illusory luminance
variation across the test stripe is weak or absent.
\labelc A real-world image  or photograph (``camera'')).
\labeld A luminance staircase, which is used to illustrate afterimages in figure \ref{Responses}c,d.}
\end{figure}%
\begin{figure}[th!]
 \centering
  \subfloat[grating induction]{\includegraphics[width=0.5\linewidth]{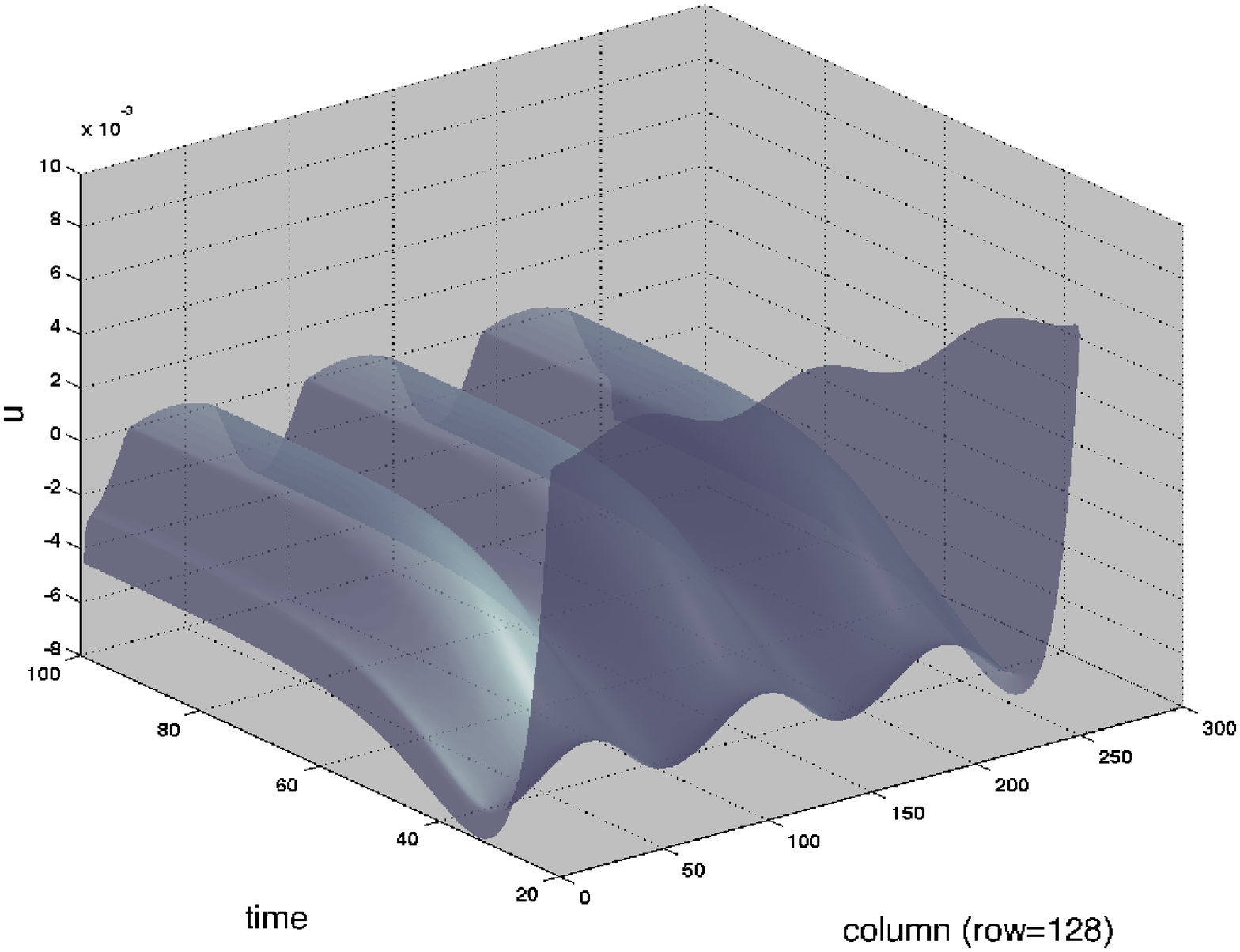}}
  \hspace*{-1cm}\subfloat[no induction]{\includegraphics[width=0.5\linewidth]{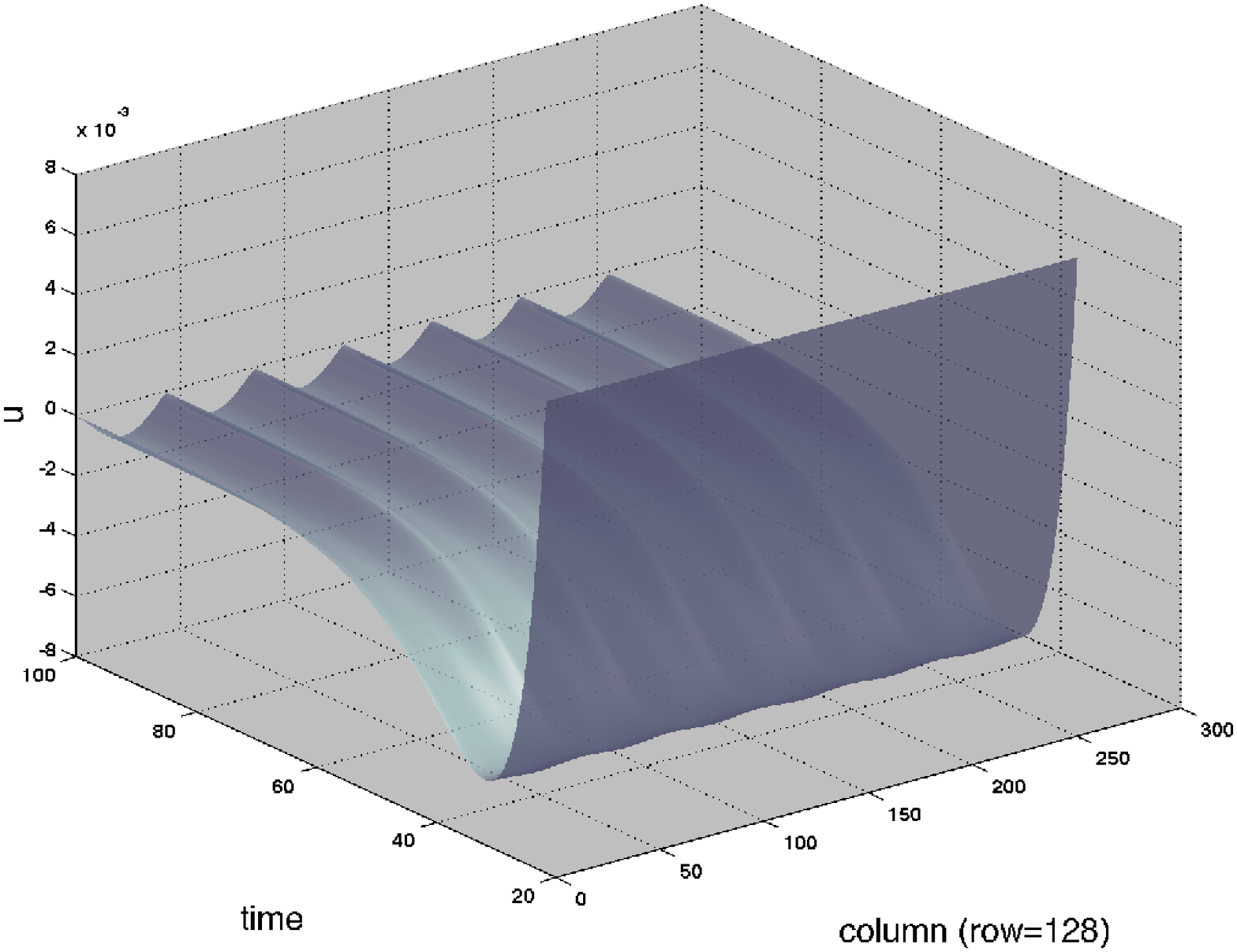}}
\mycap{GratingPrediction}{Simulation of grating induction}{Does the dynamic retina (equations \ref{retina1} \& \ref{retina2})
predict the illusory luminance variation across the test stripe ($=$ the small stripe which separates the two gratings in
figure \ref{Luminance}a,b)?
\labela Here, the image of figure \ref{Luminance}a was assigned to $I_t$.
The plot shows the temporal evolution of the horizontal line centered at the test stripe, that is it shows 
all columns $1\leq x \leq 256$ of $u_t(x,y_0)$ for the fixed row number $y_0=128$ at different instances in time $t$.
Time increases towards the background.   If values $u_t(x,y_0)>0$ ($=$ ON-responses) are interpreted as brightness,
and values $u_t(x,y_0)<0$ ($=$ OFF-responses) as darkness, then the wave pattern adequately predicts the grating
induction effect.
\labelb If the image of figure \ref{Luminance}b is assigned to $I_t$ (where human observers usually do not perceive grating
induction), then the wave-like pattern will have twice the frequency of the inducer gratings, and moreover a strongly reduced amplitude.
Thus, the dynamic retina  correctly predicts a greatly reduced brightness (and darkness) modulation across the test stripe.}
\end{figure}
\begin{figure}[th!]
 \centering
  \subfloat[ON-response (camera)]{\includegraphics[width=0.23\linewidth]{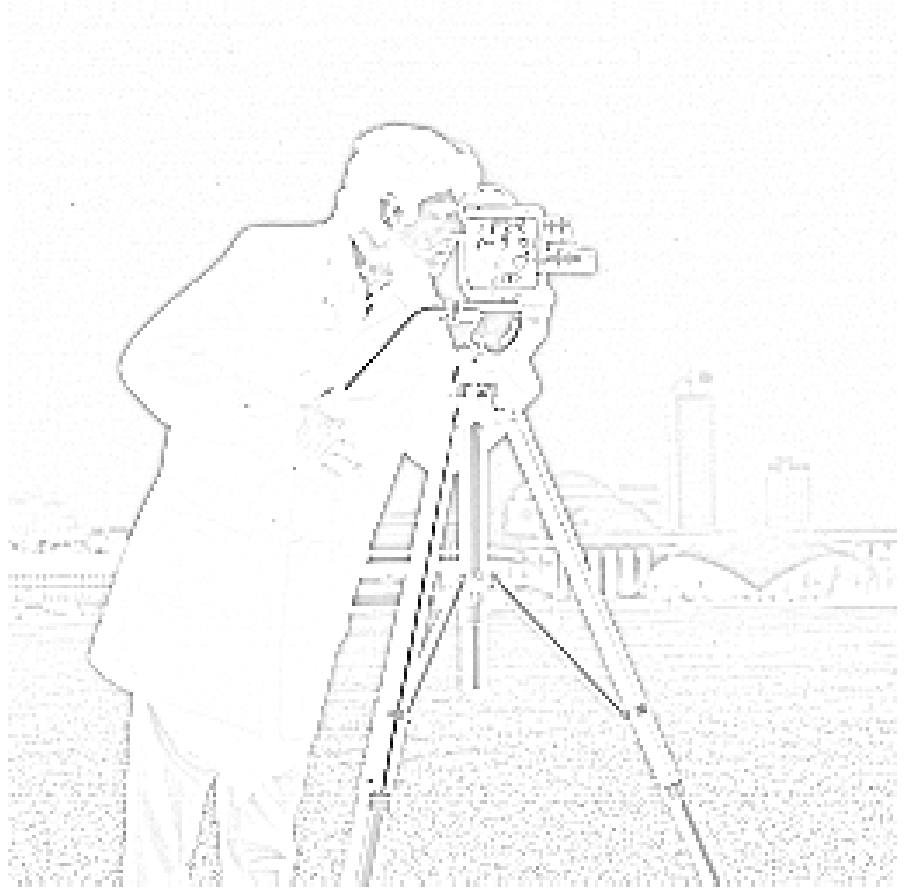}}~~
  \subfloat[OFF-response (camera)]{\includegraphics[width=0.23\linewidth]{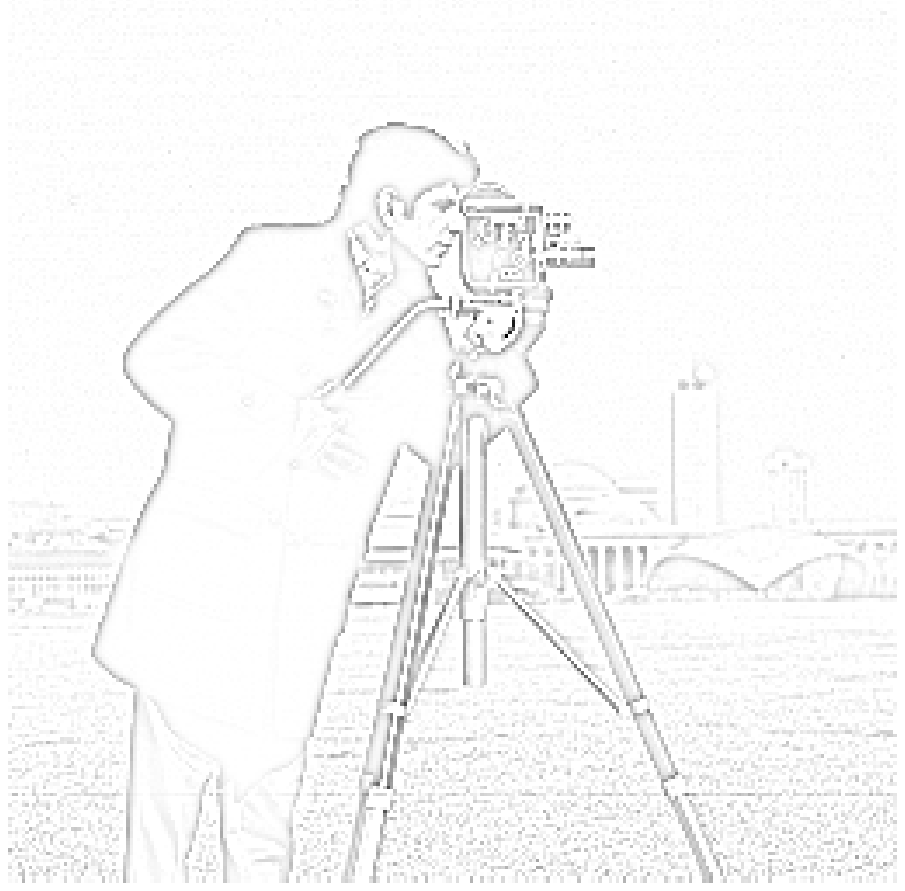}}~~
  \subfloat[ON-response (afterimage)]{\includegraphics[width=0.23\linewidth]{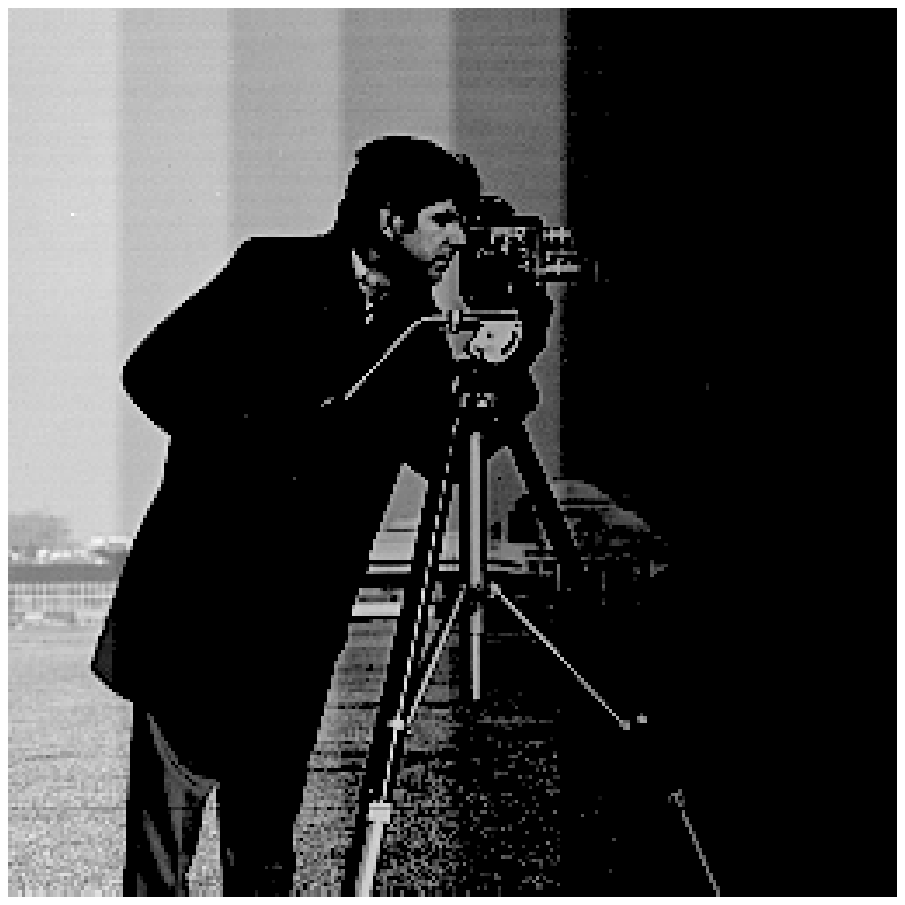}}~~
  \subfloat[OFF-response (afterimage)]{\includegraphics[width=0.23\linewidth]{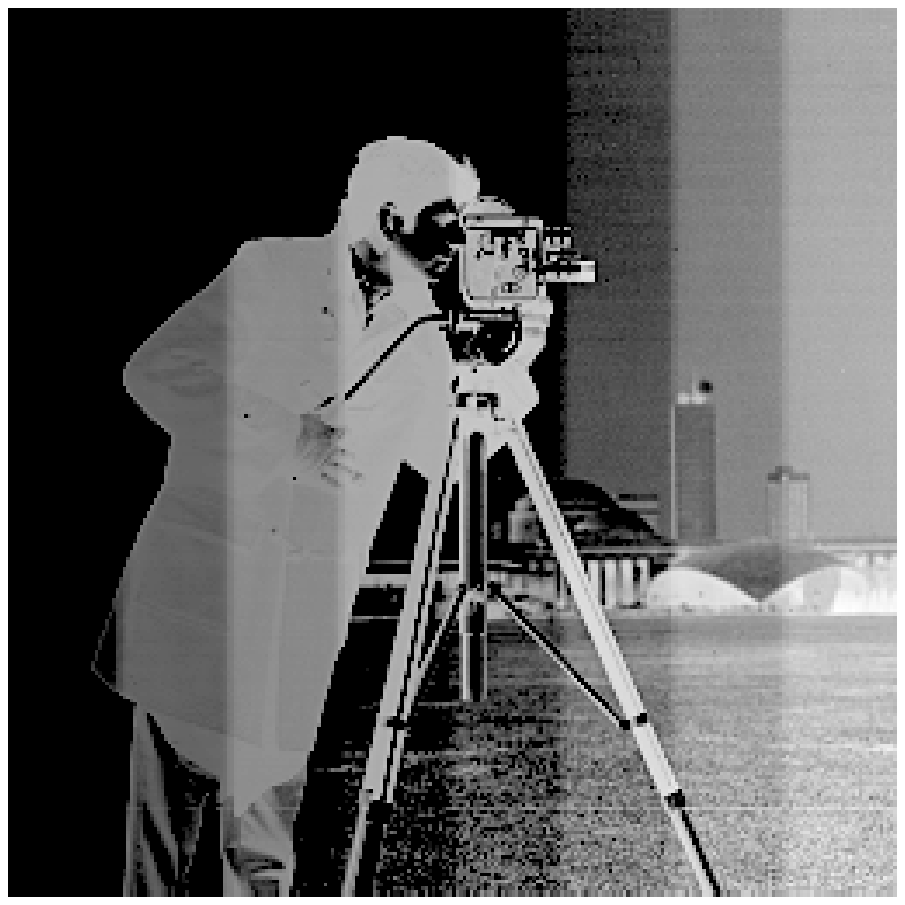}}
\mycap{Responses}{Snapshots of the dynamic retina}{%
\labela This is the ON-response (equation \ref{retina1} $[u_t(x,y)]^+$) after $t=200$ iterations of the dynamic retina
(equations \ref{retina1} \& \ref{retina2}), where the image of figure \ref{Luminance}c was assigned to $I_t(x,y)$.
Darker gray levels indicate higher values of  $[u_t(x,y)]^+$).
\labelb Here the corresponding OFF-responses ($[-u_t(x,y)]^+$) are shown, where darker gray levels indicate higher 
OFF-responses.
\labelc Until $t=200$, the luminance staircase (figure \ref{Luminance}d) was assigned to $I_t(x,y)$.  Then the image was
replaced by the image of the camera man.   This simulates a retinal saccade.  As a consequence, a ghost image of the luminance
staircase is visible in both ON- and OFF-responses (approximately until $t=250$).  From $t=260$ on, the ON- and OFF-responses
are indistinguishable from (a) and (b).   Here, \emph{brighter} gray levels indicate higher values of  $[u_t(x,y)]^+$.
\labeld Corresponding OFF-responses to  (c).  Again, \emph{brighter} gray levels indicate higher values of  $[-u_t(x,y)]^+$.
All simulations were performed with filter memory constants $\beta_1=0.9$, $\beta_2=0.85$, and diffusion coefficient
$D=0.25$.}
\end{figure}%
%
%
%
%
%=============================================================================
\section[retina]{Application 1: A dynamical retinal model\label{retina}}
%=============================================================================
%
The retina is a powerful computational device.   It transforms light intensities with different wavelengths -- as being captures
by cones (and rods for low-light vision) -- into an \emph{efficient} representation which is sent to the brain by the axons of
the retinal ganglion cells \cite{PitkovMeister12}.   The term ``efficient''  refers to \emph{redundancy reduction} (``decorrelation'') in
the stimulus on the one hand, and coding efficiency at the level of ganglion cells on the other.   Decorrelation means that 
predictable intensity levels in time and space are suppressed in the responses of ganglion cells \cite{HosoyaEtAl05,DoiEtAl2012}.
For example, a digital photograph of a clear blue  sky has a lot of spatial redundancy, because if we select a blue pixel, it is highly
probable that its neighbours are blue pixels as well \cite{Attneave1954}.  \emph{Coding efficiency} is linked to metabolic energy consumption.
Energy consumption increases faster than information transmission capacity, and organisms therefore seem to have evolved to
a trade-off between increasing their evolutionary fitness and saving energy \cite{EnergyInfoTradeOff2007}.  Retinal ganglion cells
show efficient coding in the sense that noisy or energetically expensive coding symbols are less ``used'' \cite{BalasubramanianBerry2000,PitkovMeister12}. 
Often, the spatial aspects of visual information processing by the retina are grossly approximated  by employing the \emph{Difference-of-Gaussian}
(``DoG'') model (one Gaussian is slightly broader than the other) \cite{Rodieck65}.   The Gaussians are typically two-dimensional, isotropic,
and centered at identical spatial coordinates.   The resulting DoG model is a convolution kernel with positive values in the center surrounded
by negative values.  In mathematical terms, the DoG-kernel a filter that takes the second derivative of an image.  In signal processing terms,
it is a bandpass filter (or a highpass filter if a small $3\times 3$ kernel is used).  In this way the \emph{center-surround antagonism} of (foveal) retinal
ganglion cells can be modelled \cite{Kuffler53}: ON-center ganglion cells respond when the center is more illuminated than the surround (i.e.,
positive values after convolving the DoG kernel with an image).  OFF-cells respond when the surround receives more light intensity than the
center (i.e., negative values  after convolution).   The DoG model thus assumes symmetric ON- and OFF-responses, what is again a simplification:
Differences between biological ON- and OFF ganglion cells include receptive field size, response kinetics, nonlinearities, and light-dark adaptation
\cite{ChichilniskyKalmar02,KaplanBenardete01,SymmetryBreakdown2010}.   Naturally, convolving an image with a DoG filter neither cannot
account for adaptation nor dynamical aspects of retinal information processing.  On the other hand, however, many retinal models which
target the explanation of physiological or psychophysical data are not suitable for image processing tasks.  So, biologically-inspired image
processing means that the model should solve an image processing task (e.g., boundary extraction, dynamic range reduction), while at the
same time produce predictions or has features which are by and large consistent with psychophysics and biology (e.g. brightness illusions,
kernels mimicking receptive fields, etc.).   In this spirit we now introduce a simple dynamical model for retinal processing, which re-produces
some interesting brightness illusions an could even account for afterimages.  An afterimage is an illusory percept where one continues to see
a stimulus which is physically not present any more (e.g. a spot after looking into a bright light source).  Unfortunately, the author was unable
to find a version of the model which could be strictly based on equation \ref{MembraneEquation2}.  Instead of that, here is an even more simpler
version that is based on the temporal low-pass filter (equation \ref{LPFilterPSP}).  Let $I_t(x,y)$ be a gray level image with luminance values between
zero (dark) and one (white).  The number of iterations is denoted by $t$ ($=$ discrete time).  Then:
%
		%alpha1	= 0.90;		alpha2	= 0.75;		D=0.5;		% here Hering comes out
%		alpha1	= 0.95;		alpha2	= 0.85;		D=0.1;		% no Hering

%			u	= u.*alpha1 + (1-alpha1).*(x-v);
%			v	= v.*alpha2 + (1-alpha2).*(fire(u)+x) + D*qDiffLAP(v);

%................................................................\begin{subequations}\label{psi}
\begin{eqnarray}
		u_{t+1}	&=&	\beta_1 u_t + (1-\beta_1)(I_t-v_t)																\label{retina1}\\	
		v_{t+1}	&=&	\beta_2 v_t + (1-\beta_2)([u_t]^+ +I_t) + D\cdot\vec{\nabla}^2 v_t			\label{retina2}
\end{eqnarray}
%................................................................
%
where $[u_t(x,y)]^+=$ ON-cell responses, $[-u_t(x,y)]^+=$ OFF-cell responses, $D=\mathrm{const}$ is the diffusion coefficient,
$\vec{\nabla}^2 v_t(x,y)\equiv\mathrm{div}(\mathrm{grad(v_t))}$ is the diffusion operator (Laplacian), which was discretized as  a
$3\times 3$ convolution kernel with $-1$ in the center, and $0.25$ in north, east, south, and west pixels.  Corner pixels were zero.
Thus, whereas the receptive field center is just one pixel, the surround is dynamically constructed by diffusion.  Diffusion length (and
thus surround size) depends on the filter memory constant $\beta_2$ and the diffusion coefficient $D$ (bigger values will produce a
larger surround area).\\
Figure \ref{Luminance} shows four test images which were used for testing the dynamic retina model.  The image \ref{Luminance}a
shows a visual illusion (``grating induction''), where observers perceive an illusory modulation of luminance between the two gratings,
although luminance is actually constant.   Figure \ref{GratingPrediction}a shows that the dynamic retina predicts a wave-like activity pattern
between the grating via $u_t(x,y)$.  Because ON-responses represent brightness ($=$ perceived luminance) and
OFF-responses represent darkness ($=$ perceived inverse luminance) , the dynamic retina correctly predicts grating induction. 
Does it also account for the absence of grating induction in figure \ref{Luminance}b?  The corresponding simulation is shown in
figure \ref{GratingPrediction}b, where the amplitude of the wave-like pattern is strongly reduced, and the frequency has doubled.
Thus, the absence of grating induction is adequately predicted.\\
In its equilibrium state, the dynamic retina performs contrast enhancement or boundary detection, respectively.  This is illustrated with
the ON- and OFF-responses (figure \ref{Responses}a and \ref{Responses}b, respectively) to the image of figure \ref{Luminance}c.
In comparison to an ordinary DoG-filter, however, the responses of the dynamic retina are asymmetric, with somewhat higher OFF-responses
to a luminance step (not shown).   A nice feature of the dynamic retina is the prediction of after images.  This is illustrated by
computing first the responses to a luminance staircase (figure \ref{Luminance}d), and then replacing the staircase image by
the image of the camera man (figure \ref{Luminance}c).  Figure \ref{Responses}c,d shows corresponding responses immediately
after the images were swapped.  Although the camera man image is now assigned to $I_t$, a slightly blurred afterimage of the
staircase still appears in $u_t$.  The persistence of the afterimage depends on the luminance values: Higher intensities in
the first image, and lower intensities in the second image will promote a prolonged effect.
% receptive fields in time and space --> convolutions
% learning and adaptation: temporal LP, Hebbian
% Example of image processing: the R-D retina
%
%
% barbara512_combinedOCs.eps  barbara512.eps  barbara512_retina.eps  barbara512_texture.eps
%
\begin{figure}[th!]
 \centering
  \subfloat[input ]{\includegraphics[width=0.23\linewidth]{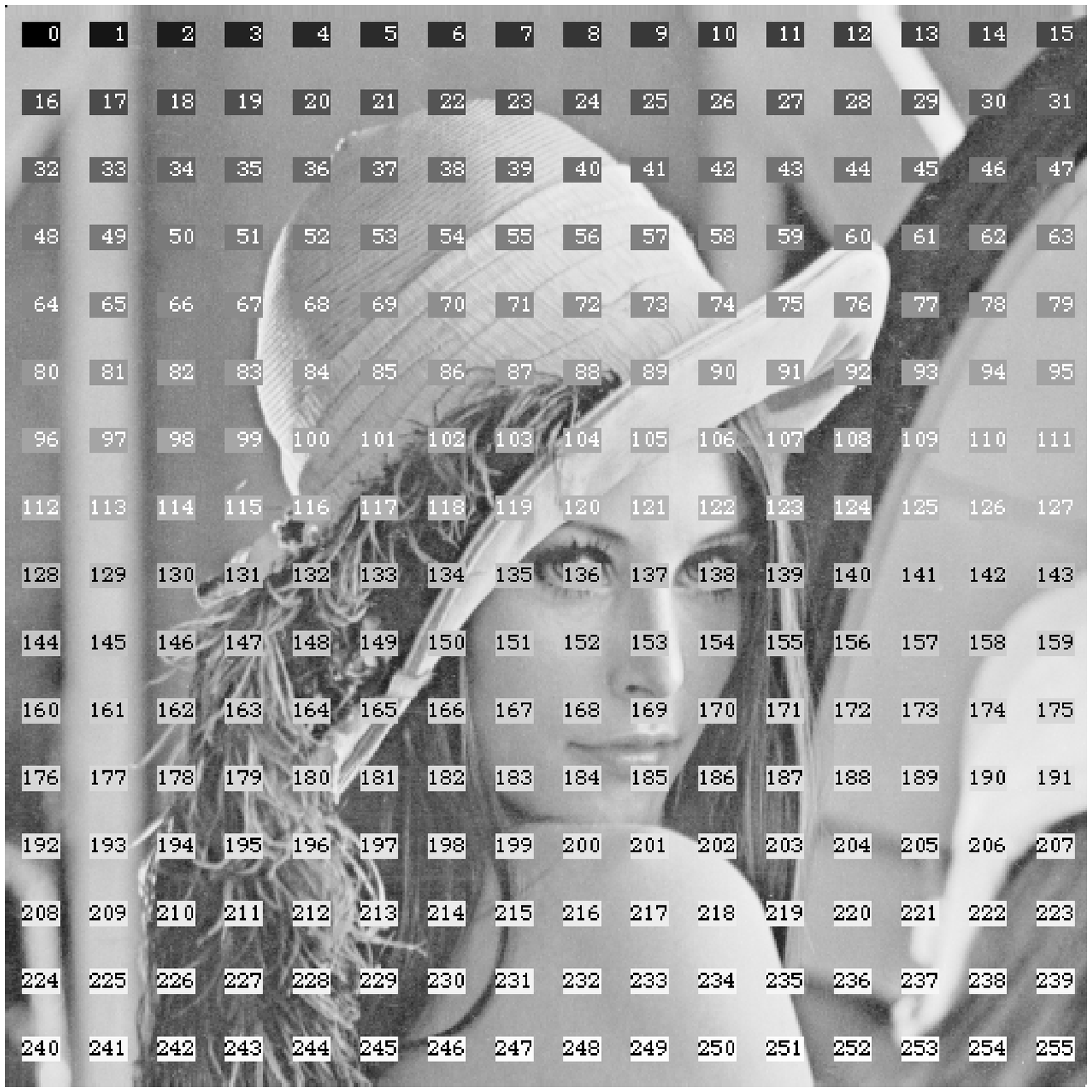}}~~
  \subfloat[retina]{\includegraphics[width=0.23\linewidth]{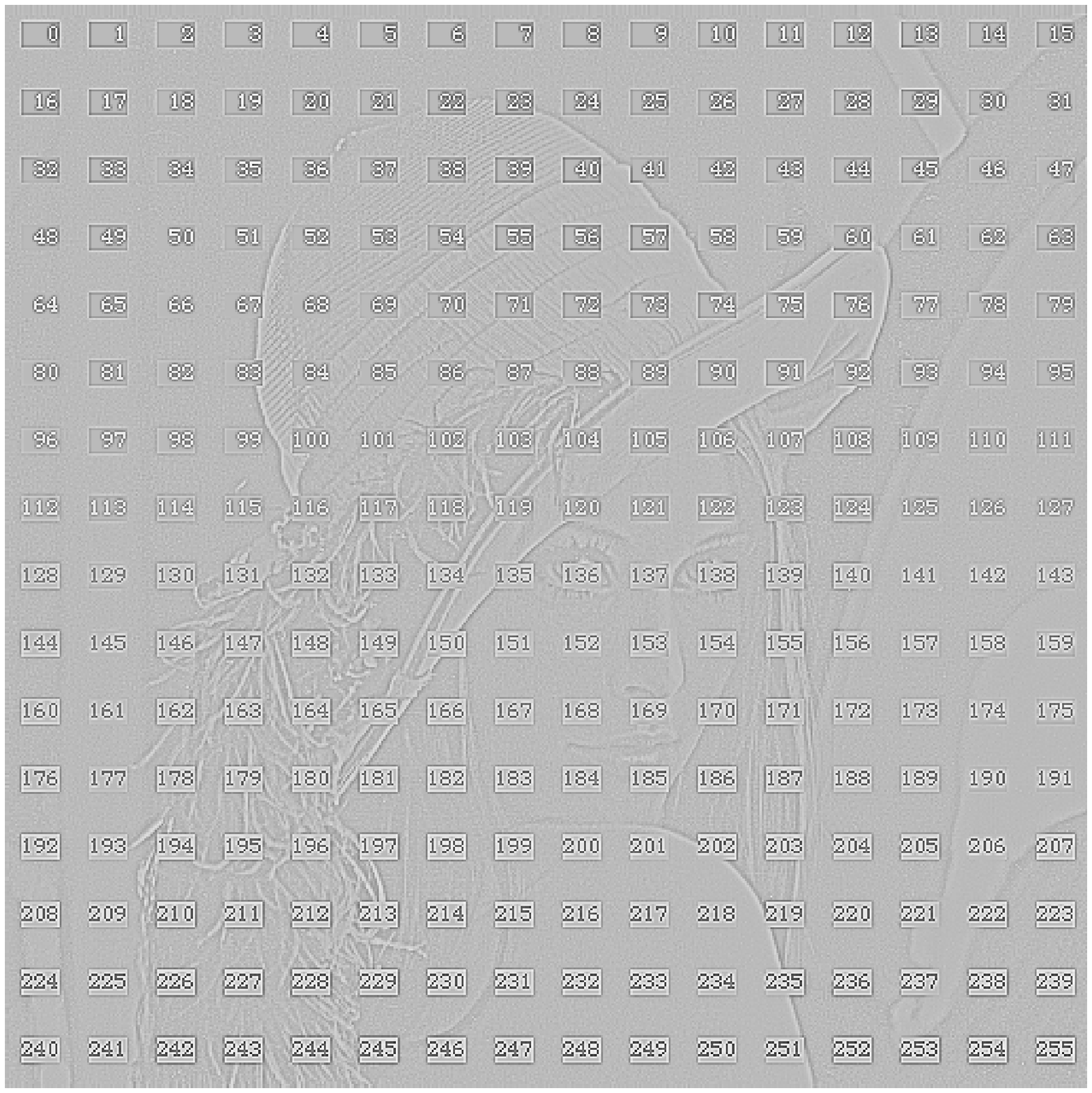}}~~
  \subfloat[$\sum$orientations]{\includegraphics[width=0.23\linewidth]{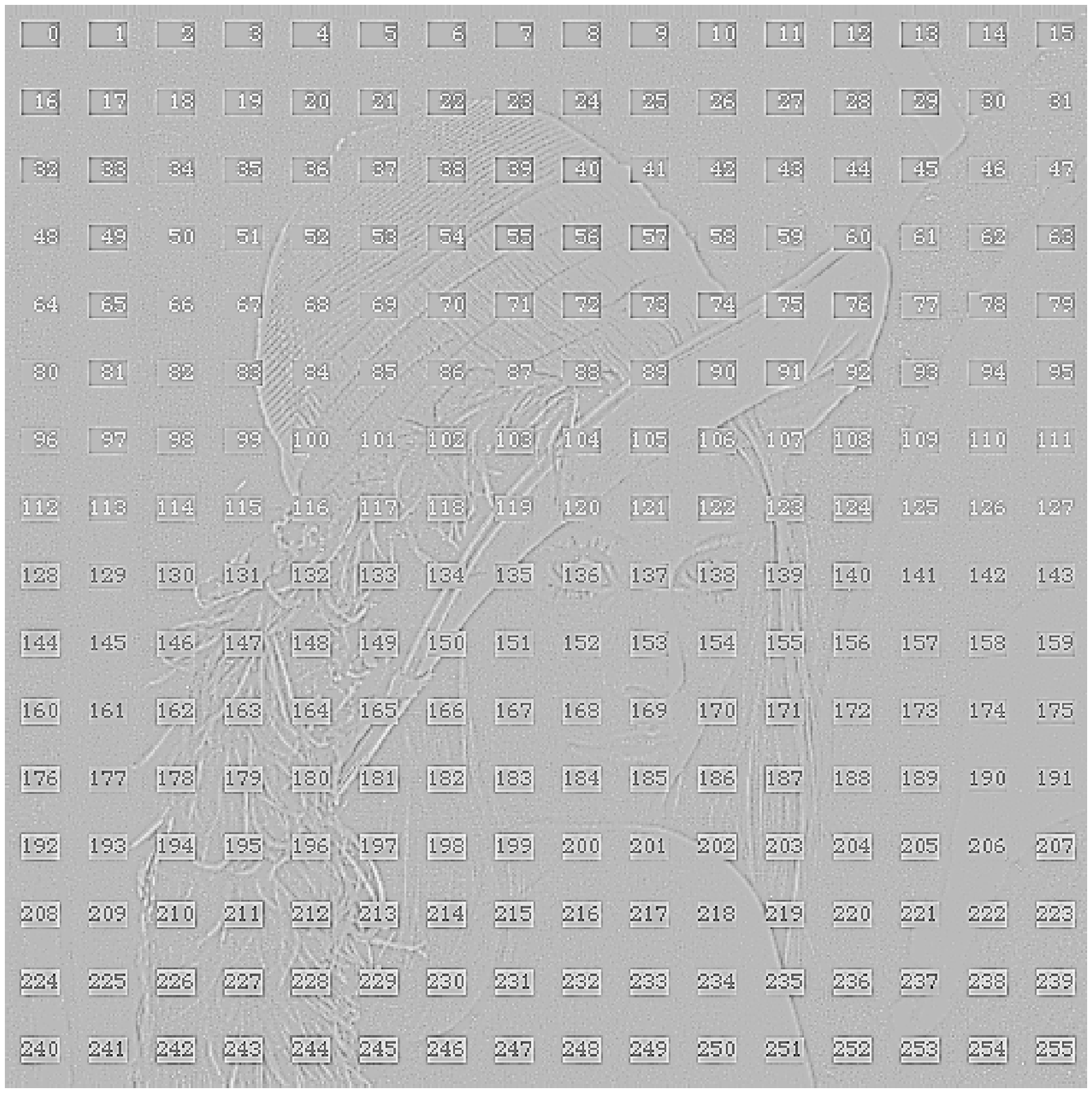}}~~
  \subfloat[texture]{\includegraphics[width=0.23\linewidth]{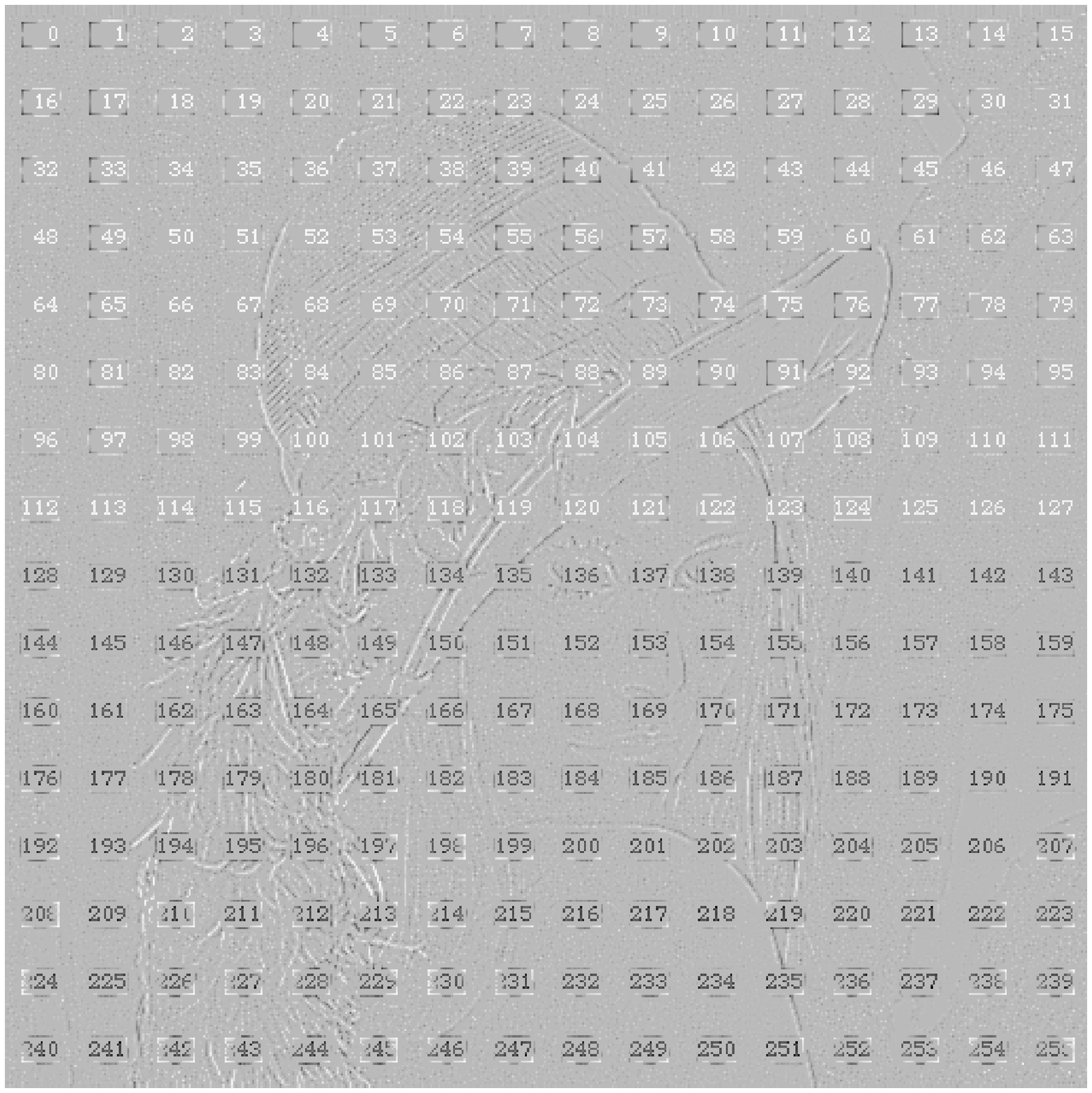}}
\mycap{Texture}{Texture segregation}{Illustration of processing a gray-scale image with the texture system.
%All processed images were contrast enhanced according to the sigmoid $\tanh[5(t_\mathrm{on}-t_\mathrm{off})]$,
%where $t_\mathrm{on}$ (texture brightness) and $t_\mathrm{off}$ (texture darkness) represent the output of the
%texture system at the respective stage (i.e., retina, $\sum$orientations, texture).
\labela Input image ``Lena" with $512\times 512$ pixels and superimposed numbers.
\labelb The output of the retina (equation \ref{SelfInhRetina}).  ON-activity is white, while OFF is black.
\labelc The analysis of the retinal image proceeds along four orientation channels. The image shows an intermediate
result after summing across the four orientations (texture brightness in white, texture darkness in black). Afterwards,
a local WTA-competition suppresses residual features that are not desired, leading to the texture representation.
\labeld This is the texture representation of the input image and represents the final output of the texture system.
As before, texture brightness is white, texture darkness is black.}
\end{figure}%
%
%
%
%
%=============================================================================
\section[texture]{Application 2: Texture segregation\label{texture}}
%=============================================================================
%
Networks based on equation \ref{MembraneEquation2} can be used for biologically plausible image processing tasks.
Nevertheless, the specific task that one wishes to achieve may imply modifications of equation \ref{MembraneEquation2}.
As an example we outline a corresponding network for segregating texture from gray scale image (``texture system''). 
We omit many mathematical details at this point (cf. chapter 4 in \cite{MatsThesis}), because they probably would make
reading too cumbersome. \\
The texture system forms part of a theory for explaining early visual information processing (\cite{MatsThesis}).
The essential proposal is that simple cells in V1 (primary visual cortex) segregate the visual input into texture, surfaces,
and (slowly varying) luminance gradients.  This idea emerges quite naturally from considering how the symmetry and scale
of simple cells relate to features in the visual world:  Simple cells with small and odd-symmetric receptive fields (RFs) 
respond preferably to contours that are caused by changes in material properties of objects such as reflectance:
Object surfaces are delimited by odd-symmetric contours.   Likewise, even-symmetrical simple cells respond 
particularly well to lines and points, which we call texture in this context.  Texture features are often superimposed on
object surfaces.  Texture features may be variable and rather be irrelevant to recognize a certain object (e.g. if the object
is covered by grains of sand), but also may correspond to an identifying feature (e.g. tree bark).
Finally, simple cells at coarse resolutions  (i.e, those with big RFs of both symmetries) are supposed to detect shallow
luminance gradients.  Luminance gradients are a pictorial depth cue for resolving the three-dimensional
layout of a visual scene.  However, they should be ignored for determining the material properties of object surfaces.\\
The first computational step in each of the three systems consists in detecting the respective features.  Following feature
detection, representations of surfaces \cite{MatsEtAl05}, gradients \cite{gradVisRes06,MatsNC06,luminosityVisRes07},
and texture \cite{MatsThesis}, respectively,  are eventually build by each corresponding  system.  Our normal visual
perception would then be the result of superimposing all three representations (brightness perception).    Having
three separate representations (instead of merely a single one) has the advantage that higher-level cortical information
processing circuits could selectively suppress or reactivate texture and/or gradient representations in addition to surface
representations.  This flexibility allows for the different requirements for deriving the material properties of objects.
For instance, surface representations are directly linked to the perception of reflectance (``lightness'') and object recognition,
respectively.  In contrast, the computation of surface curvature and the interpretation of the three-dimensional scene structure 
relies on gradients (e.g., shape from shading) and/or texture representations (texture compression with distance).\\
How do we identify texture features? We start with processing a gray scale image with a retinal model which is
based on a modification of equation \ref{MembraneEquation2}:
\begin{equation}\label{SelfInhRetina}
   \frac{dV_{ij}(t)}{dt} = \gleak(\Vrest-V_{ij}) + \zeta(\mathcal{E}_{ij}  -\mathcal{I}_{ij}) + g_{ij,si}\cdot (E_{si}-V_{ij})
\end{equation}

\def\rect#1{[ #1 ]^+}
$\mathcal{E}_{ij}$ is just the input image itself - so the center kernel (aka receptive field) is just one pixel.
$\mathcal{I}_{ij}$ is the result of convolving the input image with a $3\times 3$ surround kernel that has $1/4$ at
north, east, west and south position.  Elsewhere it is zero.  Thus, $\mathcal{E}_{ij}  -\mathcal{I}_{ij}$ approximates
the (negative) second spatial derivative of the image.  Accordingly we can define two types of ganglion cell
responses:  ON-cells are defined by $\zeta=1$  and respond preferably to (spatial) increments in luminance.
OFF-cells have $\zeta=-1$ and prefer decrements in luminance.
Figure \ref{Texture}b shows the output of equation \ref{SelfInhRetina}, which is the half-wave rectified membrane potential
$\rect{V_{ij}}\equiv\max(V_{ij},0)$.  Biological ganglion cell responses saturate with increasing contrast.  This is modeled here
by $g_{ij,si}\cdot (E_{si}-V_{ij})$, where $g_{ij,si}$ corresponds to self-inhibition - the more center and surround are activated,
the stronger.  Mathematically, $g_{ij,si} = \xi \cdot \rect{\zeta(\mathcal{E}_{ij}  -\mathcal{I}_{ij})}$, with a constant $\xi>0$
that determines how fast the responses saturate.  Why did we not simply set $g_{exc}=\mathcal{E}_{ij}$ (and $\Vexc=1$), and
$g_{inh}=\mathcal{I}_{ij}$ (and $\Vinh=-1$)  in equation \ref{MembraneEquation2} (vice versa for an OFF-cell)? Because in the
latter case ON and OFF response amplitudes would be different for a luminance step (cf. equation \ref{MembraneEquationSteadyState}).
Luminance steps are odd-symmetric features and are not texture.  Thus, we want to suppress them, and suppression is easier
if ON and OFF response amplitudes (to odd-symmetric features) are equal.\\
In response to even-symmetric features (i.e., texture features) we can distinguish two (quasi one-dimensional) response patterns.
A black line on a white background produces an ON-OFF-ON ($=$LDL) response: A central OFF-response, and two flanking OFF-responses
with much smaller amplitudes.  Analogously, a bright line on a dark background will trigger a OFF-ON-OFF or DLD response pattern.
ON- and OFF responses to lines (and edges) vary essentially in one dimension.  Accordingly, we analyze them along
four orientations.  Orientation selective responses (without blurring!) are established by convolving the OFF-channel
with an oriented Gaussian kernel, and subtract it from the (not blurred) ON-channel.  This defines texture brightness.
The channel for texture darkness is defined by subtracting blurred ON responses from OFF responses.  Subsequently,
even-symmetric response patterns are further enhanced with respect to surface features.  In order to boost
a DLD pattern, the left-OFF is multiplied with its central-ON and its right-OFF (analogous for LDL patterns).  
Note that surface response pattern (LD or DL, respectively) ideally has only one flanking response.
Dendritic trees are a plausible neurophysiological candidate for implementing such a ``logic'' AND gate
(simultaneous left \emph{and} central \emph{and} right response) \cite{LondonHaeusserl05}.\\
In the subsequent stage the orientated texture responses are summed across orientations, leaving non-oriented responses
(Figure \ref{Texture}c).
By means of a winner-takes-all (WTA) competition between adjacent (non-oriented) texture brightness (``L'')
and texture darkness (``D'') it is now possible to suppress the residual surface features on the one hand, and the flanking
responses from the texture features on the other.  For example, a LDL response pattern will generate a competition between
L and D (on the left side) and D and L (right side). Since for a texture feature the central response is bigger, it will
survive the competition with the flanking responses.  The flanking responses, however, will not.  A surface response
(say DL) will not survive either, because D- and L-responses have equal amplitudes.  The local spatial WTA-competition is
established with a nonlinear diffusion paradigm \cite{MatsPhysicaD08}.  The final output of the texture system (i.e., a
texture representation) is computed according to equation \ref{MembraneEquation2}, where texture brightness acts excitatory, 
and texture darkness inhibitory. An illustration of a texture representation is shown in figure \ref{Texture}d.
\def\ddt#1{\frac{d#1}{dt}}
\def\on#1{ #1 ^\circ}
\def\off#1{ #1 ^\bullet}
\def\rected#1{ \tilde{#1}}
\begin{figure}[th!]
 \centering
  \includegraphics[width=1.0\linewidth]{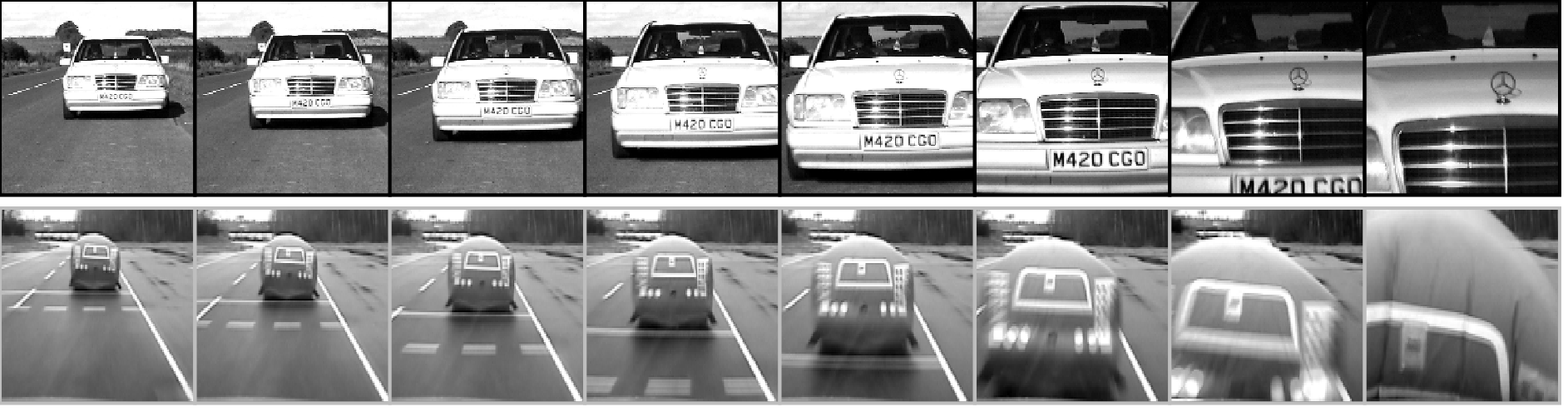}
 \mycap{VideoFrames}{Video sequences showing object approaches}{%
 Two video sequences which served as input $I_t(x,y)$ to equation \ref{advancedphoto}.
 The \textbf{top row} shows a couple of representative frames of a video where a car drives to a still observer.  Except of some camera
 shake, there is no background motion present in this video.  The car does actually not collide with the observer.
 In the video frames shown in the \textbf{bottom row}, a car (representing the observer) drives against a static obstacle.  This sequence
 implies background motion.  Here the observer actually collides with the balloon car, which flies through the air after the impact.}
\end{figure}%
\begin{figure}[th!]
 \centering
  \subfloat[a car approaches a static observer]{\includegraphics[width=0.5\linewidth]{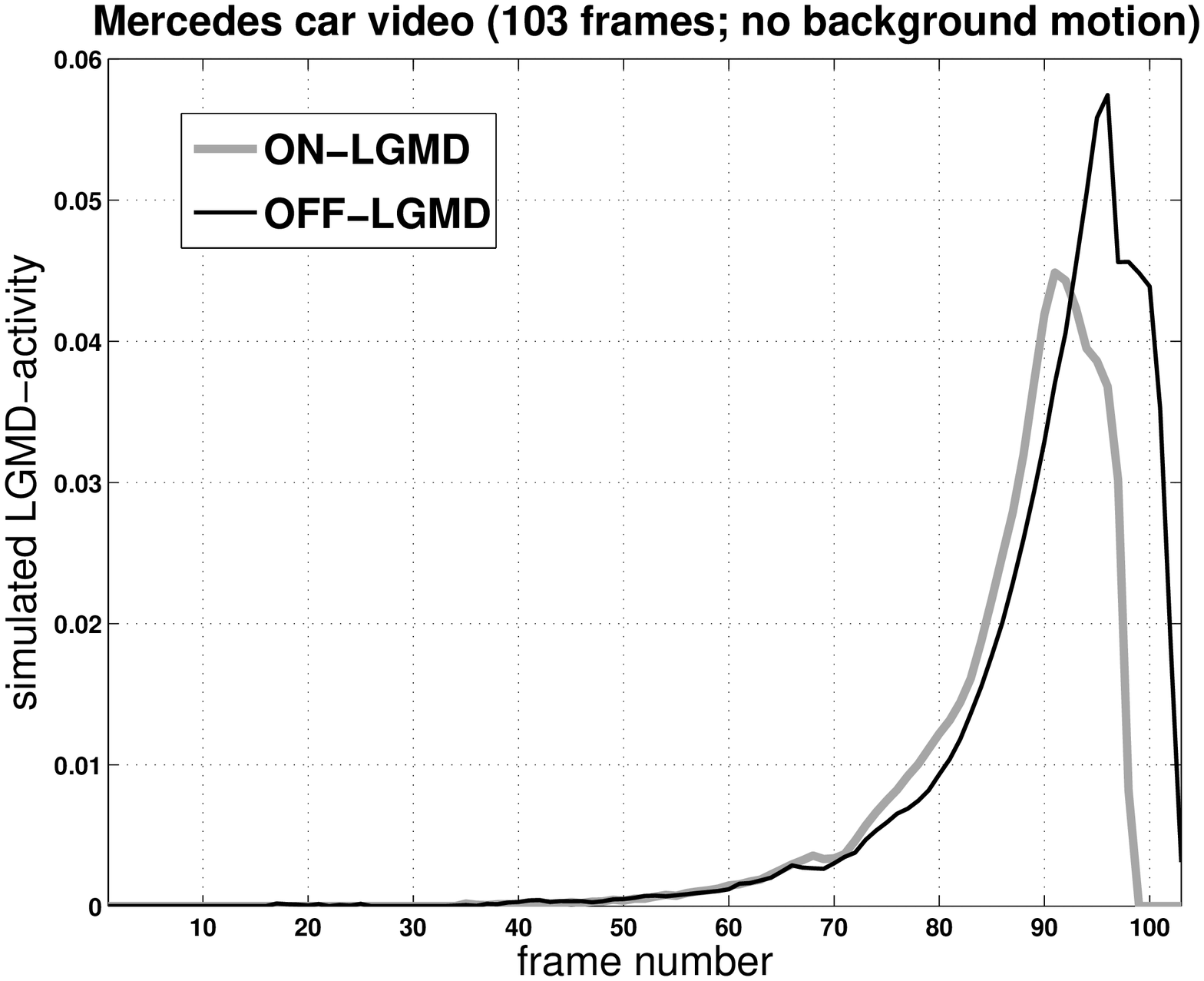}}
  \subfloat[collision with a static balloon car]{\includegraphics[width=0.5\linewidth]{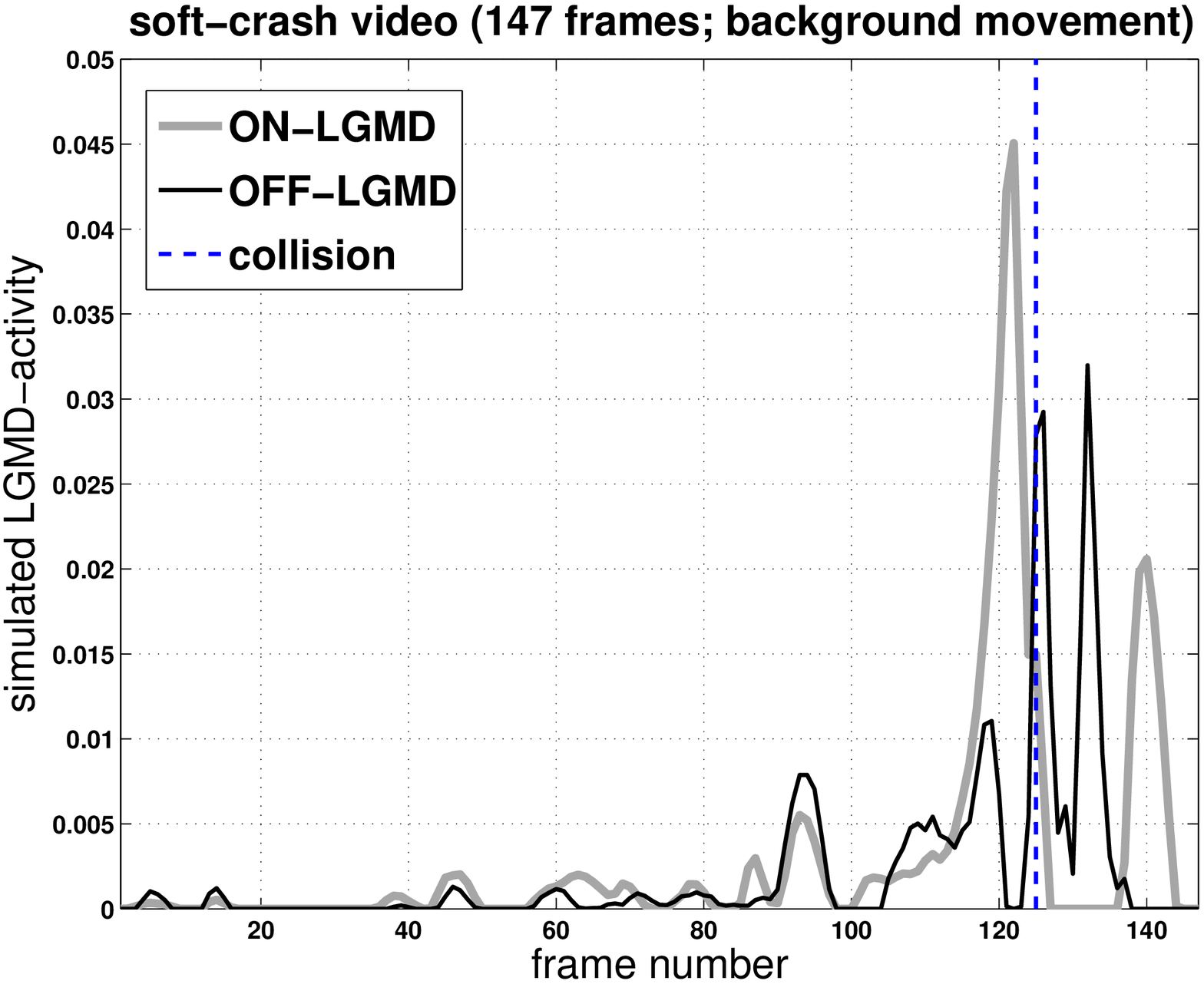}}
\mycap{LGMD}{Simulated LDMD responses}{Both figures show the rectified LGMD activities $\on{\rected{l}}_t$ (gray curves;
label ``ON-LGMD") and $\off{\rected{l}}_t$ (black curves; label ``OFF-LGMD") as computed by equation \ref{LGMDONOFF1}.
LGMD activities are one-dimensional signals that vary with time $t$ (the abscissa shows the frame number instead of time).
\labela Responses to the video shown in the top row of figure \ref{VideoFrames}.  The observer does not move and no
background motion is generated.  Both LDMD responses peak before collision would occur. 
\labelb Responses to the video shown in the bottom row of figure \ref{VideoFrames}.  Here the observer moves, and the
resulting background motion causes spurious LGMD activity with small amplitude before collision.  The collision time
is indicated by the dashed vertical line.  ON-LGMD activity peaks a few frames before collision.  OFF-LGMD activity
after collision is generated by the balloon car swirling through the air while it is moving away from the observer.}
\end{figure}%
%
%
%
%
%====================================================================================
\section[collision]{Application 3: Detection of collision threats\label{collision}}
%====================================================================================
%
%
Many animals show avoidance reactions in response to rapidly approaching objects or other animals \cite{RindSimmons99,RindSanter04}.   
Visual collision detection also has attracted attention from engineering because of its prospective applications, for example in robotics
or in driver assistant systems.  It is widely accepted that visual collision detection in biology is mainly  based on two \emph{angular variables}:
\emph{(i)} The \emph{angular size} $\Theta(t)$ of an approaching object,  and \emph{(ii)} its \emph{angular velocity} or rate of expansion
$\dot{\Theta}(t)$ (the dot denotes derivative in time).   If an object approaches an observer with constant velocity, then both angular
variables show a nearly exponential increase with time.   Biological collision avoidance does not stop here, but computes mathematical
functions (here referred to as \emph{optical variables}) of  $\Theta(t)$ and $\dot{\Theta}(t)$.  Accordingly,  three principal
classes of collision-sensitive neurons have been identified \cite{SunFrost98}.   These classes of neurons can be found in animals
as different as insects or birds.   Therefore, evolution came up with similar computational principles that are shared across many
species \cite{MatsJoan2012}.\\
A particularly well studied neuron is the \emph{Lobula Giant Movement Detector} neuron (LGMD) of the locust visual system, because
the neuron is relatively big and easy to access.   Responses of the LGMD to object approaches can be described by the so-called
\emph{eta-function}:  $\eta(t)\propto \dot{\Theta}\cdot\exp(-\alpha\Theta)$ \cite{HatsopoulosEtAl95}.   There is some evidence
that the LGMD biophysically implements $\log\eta$ \cite{GabKraKocLau02} (but see \cite{NIPS2011_0348,ECVP2012,Keil2015}): Logarithmic encoding
converts the product into a sum, with $\log\dot{\Theta}$ representing excitation and $-\alpha\Theta$ inhibition.  One distinctive
property of the eta-function is a response maximum before collision would occur.  The time of the response maximum is determined
by the constant $\alpha$ and always occurs at the fixed angular size $2\cdot\arctan(1/\alpha)$.\\
So much for the theory - but how can the angular variables be computed from a sequence of image frames?  The model which is
presented below (first proposed in \cite{MatsEliAngel04}) does not compute them explicitly, although its output resembles the eta-function. 
However, the eta-function is not explicitly computed either.  Without going too far into an ongoing debate on the biophysical details of
the computations which are carried out by the LGMD \cite{RindSimmons99,GabHatsKrapp99,NIPS2011_0348}, the model rests on lateral inhibition
in order to suppress self-motion and background movement \cite{RindBramwell96}.\\
The first stage of the model computes the difference between two consecutive image frames $I_t(x,y)$ and  $I_{t-1}(x,y)$  (assuming gray scale videos):
\begin{equation}\label{advancedphoto}
\ddt{p}  =  -\gleak \cdot p  + I_t \cdot (1-p) - I_{t-1}\cdot (1+p)
\end{equation}
where $p\equiv{p_t(x,y)}$ - we omit spatial indices $(x,y)$ and time $t$ for convenience.   The last equation and all subsequent ones derives
directly from equation  \ref{MembraneEquation2}.  Further processing in the model proceeds along two parallel pathways.  The ON-pathway
is defined by the positive values of $p$, that is $\on{p}=\max(p,0)\equiv\rect{p}$.  The OFF-pathway is defined by $\off{p}=\rect{-p}$.
In the absence of background movement, $\sum_{x,y}(\on{p}+\off{p})$ is related to angular velocity:  If an approaching object is yet far away,
then the sum will increase very slowly.  In the last phase of an approach (shortly before collision), however, the sum increases steeply.\\
The second stage are two diffusion layers $s \in \{\on{s},\off{s}\}$ (one for ON, another one for OFF) that implement lateral inhibition:
\begin{equation}
	\ddt{s}  =  \gleak (\Vrest-s)  + g_{exc}(1-s) + D\cdot\vec{\nabla}^2 s
\end{equation}
where $D$ is the diffusion coefficient (cf. equation \ref{retina2}).  The diffusion layers act inhibitory (see below) and serve to attenuate
background movement and translatory motion as caused by self-movement.   If an approaching object is sufficiently far away, then the spatial
variations (as a function of time) in $p$ are small.  Similarly, translatory motion at low speed will also generate small spatial displacements.   Activity propagation proceeds at constant
speed and thus acts as a predictor for small movement patterns.   But diffusion cannot keep up with those spatial variations as they are
generated in the late phase of an approach.   The output of the diffusion layer is $\on{\rected{s}}=\rect{\on{s}}$ and $\off{\rected{s}}$,
respectively (the tilde denotes half-wave rectified variables).\\
The input into the diffusion layers  $\on{g}_{exc}=250\cdot\on{\rected{v}}$ and $\off{g}_{exc}=250\cdot\off{\rected{v}}$, respectively, 
is brought about by feeding back activity from the third stage of the model:
\begin{equation}\label{summingUnits}
	\ddt{v}=  \gleak (\Vrest-v)  + g_{exc}(1-v) -  g_{inh}(0.25+v)
\end{equation}
with excitatory input $\on{g}_{exc}=250 \on{p}\cdot \exp(-500\on{\rected{s}})$ and inhibitory input  $\on{g}_{inh}=500 \on{\rected{s}}$
($\off{g}_{exc}$ and $\off{g}_{inh}$ analogously).   Hence, $v$ receives two types of inhibition from $\rected{s}$.  First,  $\rected{s}$
directly inhibits $v$ via the inhibitory input $g_{inh}$.  Second, $\rected{s}$ gates the excitatory input $g_{exc}$:  Activity from the
first stage is attenuated at those positions where $\rected{s}>0$.  This means that $v$ can only decrease where $\rected{s}>0$ and feedback
from $v$ to $s$ assures that also the activity in $s$ will not grow further then.  In this way it is largely avoided that the diffusion
layers continuously accumulate activity and will eventually ``drown" (i.e., $\rected{s}_t(x,y)>0$ at all positions $(x,y)$).   Drowning
otherwise would occur in the presence of strong background motion,  making the model essentially blind to object approaches.  The use
of two diffusion layers contributes to a further reduction of  drowning.\\
The fifth and final stage of the model represents the LGMD neuron and spatially sums the output from the previous stage:
\begin{equation}\label{LGMDONOFF1}
	\ddt{l} = \gleak (\Vrest-l) +	g_{exc} (1-l)
\end{equation}
where $l\in\{\on{l}, \off{l}  \}$ and $\on{g}_{exc}=\gamma\cdot \sum_{x,y}\on{\rected{v}}(x,y)$ and analogously for $\off{g}_{exc}$.
$\gamma$ is a synaptic weight.  Notice that whereas  $p,s$ and $v$ are two-dimensional variables in space, $l$ is a scalar.  The final
model output corresponds to the two half-wave rectified LGMD activities $\on{\rected{l}}_t$ and $\off{\rected{l}}_t$, respectively.
Figure \ref{VideoFrames} shows representative frames from two video sequences that were used as input to the model.   Figure
\ref{LGMD} shows the corresponding output as computed by the last equation.  Simulated LGMD responses are nice and clean in the
absence of background movement (Figure \ref{LGMD}a).  The presence of background movement, on the other hand, produces spurious
LGMD activation before collision occurs (Figure \ref{LGMD}b).
% 
%
%
%
%
%
%
%=============================================================================
%\section[networks]{Networks of neurons}
%=============================================================================
%
% Dynamic phenomena: Synchronization, oscillations, criticiality, up-down states ...
% Example models: Balanced activity (dynamics) with and w/o Hebb, STDP
% How to generate input? Poisson versus ongoing activity (compare network activity with Poisson)
% Properties: faster responses, enhanced DR
% http://www.scholarpedia.org/article/Balance_of_excitation_and_inhibition
%
%==================================================================================
\section[Discussion]{Conclusions\label{discussion}}
%==================================================================================
%
% choice of integration method: Speed versus accuracy
% spike versus rate
% dendrites or not (not treated here; pyramidal neurons can be treated as two-layer NN apical vs. basal dendrite)
% different neuron models:
% Hodgkin-Huxley
% Izhikevich http://www.scholarpedia.org/article/Bursting
% http://www.scholarpedia.org/article/Conductance-based_models
% http://www.scholarpedia.org/article/Morris-Lecar_model
% http://www.scholarpedia.org/article/FitzHugh-Nagumo_model
Neurodynamical models (which are suitable for processing real-world images) and PDE ($=$partial differential equation)-based image processing
algorithms typically differ in the  way they are designed, and in their respective predictions with regard to neuroscience.  
PDE-based image processing algorithms derive often from an optimization principle (such as minimizing an energy functional).  As a result,
a set of differential equations is usually obtained which evolves over time \cite{ChanShenVese2003}.   Many of these algorithms
are designed \emph{ad hoc} for specific image processing tasks  (e.g. optical flow computation \cite{AubertDericheKornprobst99},
segmentation \cite{Vitti2012}, or denoising \cite{RudinOsherFatemi1992}), but they are usually not designed according to neuronal circuits.
Similarly they usually do not predict psychophysical results.   Some remarkable exceptions, however, do exist.
For example, the color enhancement algorithm described in reference \cite{VarColorCorr2009} is based on a perceptually motivated  energy
functional, which includes a contrast term (enforcing local contrast enhancement) and a dispersion term (implementing  the gray-world assumption
and enforcing fidelity to the data).   In a similar vein could the Retinex algorithm - for estimating perceived reflectance ($=$lightness) -
also be casted into a variational framework \cite{VarRetinex2003}.  An algorithm for tone mapping of high dynamic range images was
originally motivated by compressing high contrasts while preserving low contrasts \cite{GradienDomainHDR2002}.  Although the authors
did not explicitly acknowledge any inspiration from neuroscience, it is nevertheless striking how the algorithm resembles
filling-in architectures \cite{MatsEtAl05}.  Filling-in has been proposed as a mechanism for computing smooth representations of
object surfaces in the visual system.   Smooth representations means that surfaces are ``tagged'' with perceptual value such as
color, movement direction, depth or lightness \cite{Komatsu2006}.  Filling-in is often modeled by (lateral) activity propagation
within compartments, which are defined through contrast boundaries \cite{GrossMing87}.  A recently proposed filling-in based
computer vision algorithm identifies regions with  a coherent movement direction from a (initially noisy) optic flow field \cite{BayerlNeumann2007}.
The algorithm proposes also a solution to the so-called aperture problem, and is based on corresponding computations of the brain's visual system
\cite{BayerlNeumann2004}.   A further method that resembles the filling-in process is image impainting (e.g., \cite{ImpaintingFramework2010}).
Image impainting completes missing regions by propagating the structure of the surround into that region.   Although image impainting has not
been related to neurophysiological principles, similar (slow) filling-in effects seem to exist also in the brain (e.g., texture filling-in \cite{Motoyoshi1999}).
%
%==================================================================================
\section*{Acknowledgements}
%==================================================================================
%
MSK acknowledges support from a \emph{Ramon \& Cajal} grant from the Spanish government, the \emph{``Retenci{\'o}n de talento''}
program from the University of Barcelona, and the national grants \emph{DPI2010-21513} and \emph{PSI2013-41568-P}. 
 {\small
\bibliography{refs}{}
\bibliographystyle%
  {bicv}
  % {wivchnum}%for numerical citation and numerically listed entries in the bibliography
  %{wivchauy}%for author--year citation and alphabetical order in the bibliography
}

%...
%\appendix
%\include{a12}
%\backmatter
%\include{b12}
%\printindex
\end{document}